\documentclass[lettersize,journal]{IEEEtran}
\usepackage{graphicx} 
\usepackage{adjustbox}
\usepackage{amsmath}
\usepackage{float}
\usepackage{subfig}
\usepackage{caption}
\usepackage{tabularx}
\usepackage[table,xcdraw]{xcolor}
\usepackage{pifont}
\usepackage{changepage}
\usepackage{gensymb}
\usepackage{fontawesome}
\usepackage{multirow} 
\usepackage[edges]{forest}
\usepackage{hyperref}
\usepackage{makecell}
\usepackage{amssymb}

\definecolor{lightgreen}{rgb}{0.9, 1, 0.9}
\definecolor{lightred}{rgb}{1, 0.9, 0.9}
\definecolor{lightblue}{rgb}{0.9, 0.9, 1}
\definecolor{lightyellow}{rgb}{1, 1, 0.8}

\usepackage{verbatim}

\usepackage[boxed,ruled,vlined,linesnumbered,commentsnumbered, noresetcount]{algorithm2e}
\usepackage{algorithmic}

\usepackage{diagbox}
\usepackage{tikz}
\usetikzlibrary{shapes.geometric}
\usetikzlibrary{positioning}

\usepackage{tabularx}
\usepackage{ragged2e} 
\newcolumntype{L}{>{\RaggedRight}X} 
\usepackage{lipsum} 
\usepackage{amsmath}
\usepackage{amssymb}

\makeatletter
\newcommand\notsotiny{\@setfontsize\notsotiny\@vipt\@viipt}
\makeatother

\begin{document}

\title{A Survey and Tutorial of Redundancy Mitigation for Vehicular Cooperative  Perception: Standards, Strategies and Open Issues}


\author{

\IEEEauthorblockN{Tengfei Lyu, Md Noor-A-Rahim, Dirk Pesch, and Aisling O'Driscoll}

\thanks{Tengfei Lyu, Md. Noor-A-Rahim,  Dirk Pesch and Aisling O'Driscoll are with the nasc research, School of Computer Science \& IT, University College Cork,  Ireland. 


(E-mail: {\tt t.lyu@cs.ucc.ie, \{md.noorarahim, dirk.pesch, aisling.odriscoll\}@ucc.ie})
}
}

\maketitle
\begin{abstract}
 This paper provides an in-depth review and discussion of the state of the art in redundancy mitigation for the vehicular Collective  Perception Service (CPS). We focus on the evolutionary differences between the 2019 redundancy mitigation rules proposed in \textit{ETSI TR 103 562} 
 versus the 2023 technical specification \textit{ETSI TS 103 324} 
 which uses a Value of Information (VoI) based mitigation approach. 
 We also critically analyse the academic literature that has sought to quantify the communication challenges posed by the CPS
 and present a unique taxonomy of the redundancy mitigation approaches proposed 
 using three distinct classifications: \textit{object inclusion filtering}, \textit{data format optimisation}, and \textit{frequency management}. 
Finally, this paper identifies open research challenges that must be adequately investigated to satisfactorily deploy CPS redundancy mitigation measures. Our critical and comprehensive evaluation serves as a point of reference for those undertaking research in this area. 
 
\end{abstract}



\section{Introduction}
The advent of connected and autonomous vehicles (CAVs) marks a transformative shift in the transportation landscape, promising enhanced safety, efficiency, and mobility. Central to the functionality of CAVs is the concept of the \textit{Collective Perception Service (CPS)}. This paradigm enables vehicles to share sensory information facilitating real-time collaboration to construct a more comprehensive understanding of their environment. This is especially important in complex and congested urban landscapes. Unlike Collective Awareness Messages (CAMs)~\cite{gunther2015potential} that provide data about the transmitting vehicle, the CPS allows vehicles to share information about objects detected in their local environment, going beyond the limitations of their own onboard sensors. The CPS enables the broadcasting of \textit{Collective Perception Messages (CPMs)}, commonly referred to as \textit{cooperative perception}, which contains data about the positions, velocities, and attributes of detected objects. This shared perception is vital for detecting non-connected road users and obstacles not within direct line of sight, such as those hidden by other vehicles or building infrastructure. This can lead to proactive safety measures, such as early collision warnings, adaptive cruise control adjustments, and better decision-making for autonomous vehicles. In particular, such cooperatively shared information can supplement the local sensor data used by autonomous vehicles to navigate complex environments with external information, providing a more comprehensive and extended understanding of their surroundings~\cite{malik2023collaborative}.

However, message transmission redundancy occurs in the CPS when multiple vehicles detect and report the same object. This redundant information transmission challenges collective perception systems and creates a high traffic load on the vehicular network. While redundancy can be beneficial in terms of enhanced robustness and reliability, excessive redundancy can have a negative impact on the communication network performance. We categorise two main forms of redundancy: \textit{spatial redundancy}, which occurs when nearby vehicles detect and report the same object within a similar time interval, and \textit{temporal redundancy}, which arises when vehicles report the same object repeatedly over short intervals without significant changes in the object's state. Excessive redundancy overloads the wireless channel, reducing its capacity to transmit new or critical information and increasing latency due to network congestion. Both issues can be critical for time-sensitive applications. Furthermore, for human-driven vehicles, unfiltered sharing of CPM object data risks overwhelming the driver's cognitive capacity, hindering timely and effective decision-making. Therefore, efficient filtering is required to ensure that only the most relevant and actionable data is presented. While this can occur on the receiver side, it is preferable that such redundant information not be transmitted if it is not relevant. 

In an attempt to tackle the redundancy challenge, ETSI has sought to specify a number of CPM redundancy mitigation rules for deciding what objects should be included in a CPM, along with measures to minimise the frequency of transmissions. This acknowledges that the default CPM rules fail to achieve a reasonable communication burden.
Initial redundancy mitigation proposals were specified in 2019 as part of ETSI \textit{TR 103 562}~\cite{etsiintelligent2019} under the Release 1 C-ITS architecture, although the CPS was not officially part of Release 1. These measures were subsequently extended and standardised in 2023 in \textit{ETSI TS 103 324}~\cite{etsiintelligent} as part of the Release 2 C-ITS architecture. In \textit{ETSI TR 103 562}, such proposals were referred to as \textit{redundancy mitigation techniques} based on \textit{Redundant Mitigation Rules (RMRs)}. In \textit{ETSI TS 103 324}~\cite{etsiintelligent}, they are instead referred to as \textit{Object Inclusion Rate Control} measures based on a \textit{Value of Information (VoI)} measure. The newer VoI-based rules employ the same techniques as specified in the older RMRs but are extended with additional filtering rules, new VoI-based measures for frequency and content management and a proposal to leverage ITS-S multi-channel operation (MCO) capabilities if available.  None of the newly specified VoI-based measures have been evaluated in the literature. 



Some survey papers have reviewed the collective perception service, but none have provided a comprehensive analysis of redundancy mitigation measures in the context of the latest ETSI CPS specification. Huang et al.~\cite{huang2023v2x} conducted a thorough review of CP frameworks and methodologies. Although the authors briefly introduced redundancy mitigation techniques, categorising them into rule-based, distance-based, and learning-based approaches, their primary emphasis was on CP frameworks rather than an in-depth examination of redundancy mitigation. Similarly, Bai et al.~\cite{bai2024survey} presented an extensive survey of CP architectures and sensor fusion schemes, outlining early, intermediate, and late fusion strategies and discussing trade-offs in communication capabilities. However, they paid limited attention to communication quality and redundancy mitigation techniques. Caillot et al.~\cite{caillot2022survey} provided a comprehensive overview of CP approaches and frameworks, emphasising the enhancement of situational awareness through multimodal sensor data fusion. While they acknowledged communication bandwidth, latency, and synchronisation as critical factors, these aspects were not subject to an in-depth discussion. In contrast, this survey provides an in-depth review of the most recent ETSI CPS specification, with a particular focus on the object inclusion rules and redundancy mitigation measures. Unlike existing literature, which is largely based on \textit{ETSI TR 103 562}, a particular emphasis is placed on the updated VoI-based mitigation measures specified in \textit{ETSI TS 103 324}~\cite{etsiintelligent} and a thorough summary and critical analysis is conducted of academic literature that has sought to quantify the communication challenges posed by the CPS and that propose or evaluate measures to address this. This includes academic literature that has evaluated the older ETSI redundancy rules but also proposed novel redundancy mitigation mechanisms. Importantly, a unique classification method is proposed, categorising redundancy mitigation strategies into three distinct classes: \textit{object inclusion filtering}, \textit{data format optimisation}, and \textit{frequency management}. This classification provides a clear framework for understanding and extending the state of the art. Finally, open research problems related to redundancy mitigation for the collective perception service are discussed that remain unresolved or under-investigated.  

The remainder of the paper is structured as follows: Section~\ref{standards} introduces the ETSI CPM specification, including the CPM message structure and generation process, and the default object inclusion rules and additional optional redundancy mitigation measures. The evolution of the specification with respect to message structure and redundancy mitigation measures is particularly emphasised.  Section~\ref{network} reviews the literature that evaluates the challenge that the collective perception service poses for vehicular communication networks and analyses academic papers that have sought to evaluate the performance of ETSI specified redundancy mitigation rules. Section~\ref{rmrs} presents the academic literature proposing alternative, non-standardised redundancy mitigation approaches. Section~\ref{eval} reviews the tools used for quantitative performance evaluation of such systems, emphasising why one might choose certain environments depending on the focus of the study. Finally, Section~\ref{orcs} presents open research problems related to redundancy mitigation for the collective perception service. A table of commonly used acronyms has been provided in Table~\ref{Abbreviations}.

\begin{table}[htbp]
\caption{Acronyms.}
\label{Abbreviations}
\centering
\scriptsize 
\begin{tabular}{|p{0.2\linewidth}|p{0.5\linewidth}|}
\hline
\multicolumn{1}{|c|}{\textbf{Abbreviations}} & \multicolumn{1}{c|}{\textbf{Full name}} \\
\hline

AI & Artificial Intelligence \\
\hline
AoI & Age of Information \\
\hline
AP & Average Precision \\
\hline
BEV & Bird's-eye View \\
\hline
BME & Bandwidth Management Entity \\
\hline
CAM & Collective Awareness Message \\
\hline
CAV & Connected Autonomous Vehicle \\
\hline
CBR & Channel Busy Ratio \\
\hline
CPA & Cooperative Perception Awareness \\
\hline
CPM & Collective Perception Message \\
\hline
CPS & Collective Perception (CP) Service \\
\hline
DCC & Decentralised Congestion Control \\
\hline
EAR & Environmental Awareness Ratio \\
\hline
FCL & Functional Configuration Limit \\
\hline
FCP & Functional Configuration Profile \\
\hline
FoV & Field of View \\
\hline
ITS & Intelligent Transport Systems \\
\hline
ITS-S & Intelligent Transport Systems Station \\
\hline
LA & Look-Ahead \\
\hline
LDM & Local Dynamic Map \\
\hline
MAPEM & MAP (topology) Extended Message \\
\hline
MCCP & Minimum Cost Coverage Problem \\
\hline
MCO & Multi-Channel Operation \\
\hline
ML & Machine Learning \\
\hline
MTU & Maximum Transmission Unit \\
\hline
MTU & Maximum Transmission Unit \\
\hline
PF & Pre-Filter \\
\hline
PDR & Packet Delivery Ratio \\
\hline
POC & Perception Object Container \\
\hline
PRC & Perception Region Container \\
\hline
RB & Resource Block \\
\hline
RL & Redundancy Level \\
\hline
RMR & Redundancy Mitigation Rule \\
\hline
RMR & Redundancy Mitigation Rule \\
\hline
RSU & Road Side Unit \\
\hline
RMSE & Root Mean Square Error \\
\hline
SEM-COM & Semantic Communication \\
\hline
SIC & Sensor Information Container \\
\hline
V2I & Vehicle-to-Infrastructure \\
\hline
V2V & Vehicle-to-Vehicle \\
\hline
V2X & Vehicle-to-Everything \\
\hline
VoI & Value of Information \\
\hline
VRU & Vulnerable Road User \\
\hline
WGS84 & World Geodetic System 1984 \\
\hline

\end{tabular}
\end{table}

\section{Standardisation of Collective Perception\label{standards}}

\begin{figure*}
    \centering
    \includegraphics[width=\textwidth]{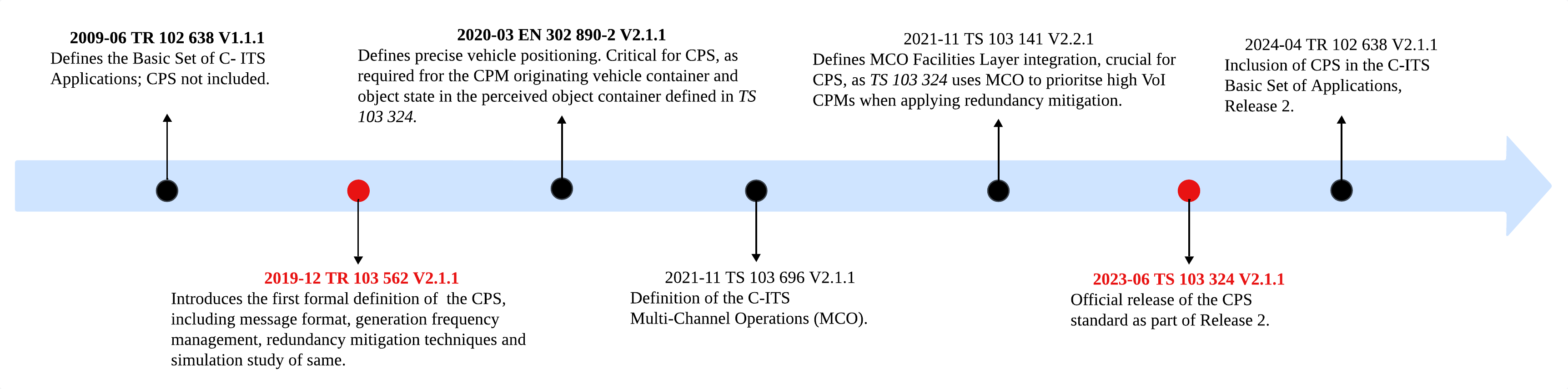}
    \caption{Timeline of the ETSI CPS standard and associated standards that inform proposed redundancy mitigation techniques.}
    \label{fig:ETSI_Timeline}
\end{figure*}

The ETSI CPS was formally specified in C-ITS Release 2 but informed by an earlier technical report as part of the Release 1 C-ITS specifications. For the remainder of this paper, we will refer to these documents as \textit{ETSI CPS TS (2023)} and \textit{ETSI CPS TR (2019)}, respectively.  Importantly, as many other standards inform the specification of the CPS, and in particular its approach to redundancy mitigation, we have shown a time-line of standards development in Fig.~\ref{fig:ETSI_Timeline}, indicating their importance to the development of the CPS. To the best of the authors' knowledge, no papers in the literature have evaluated the extensions introduced in \textit{ETSI CPS TS (2023)} or highlighted the structural changes made to the CPS in Release~2.
 
\subsection{ETSI CPM Message Structure} 
\begin{figure}
    \centering
    \includegraphics[width=\columnwidth]{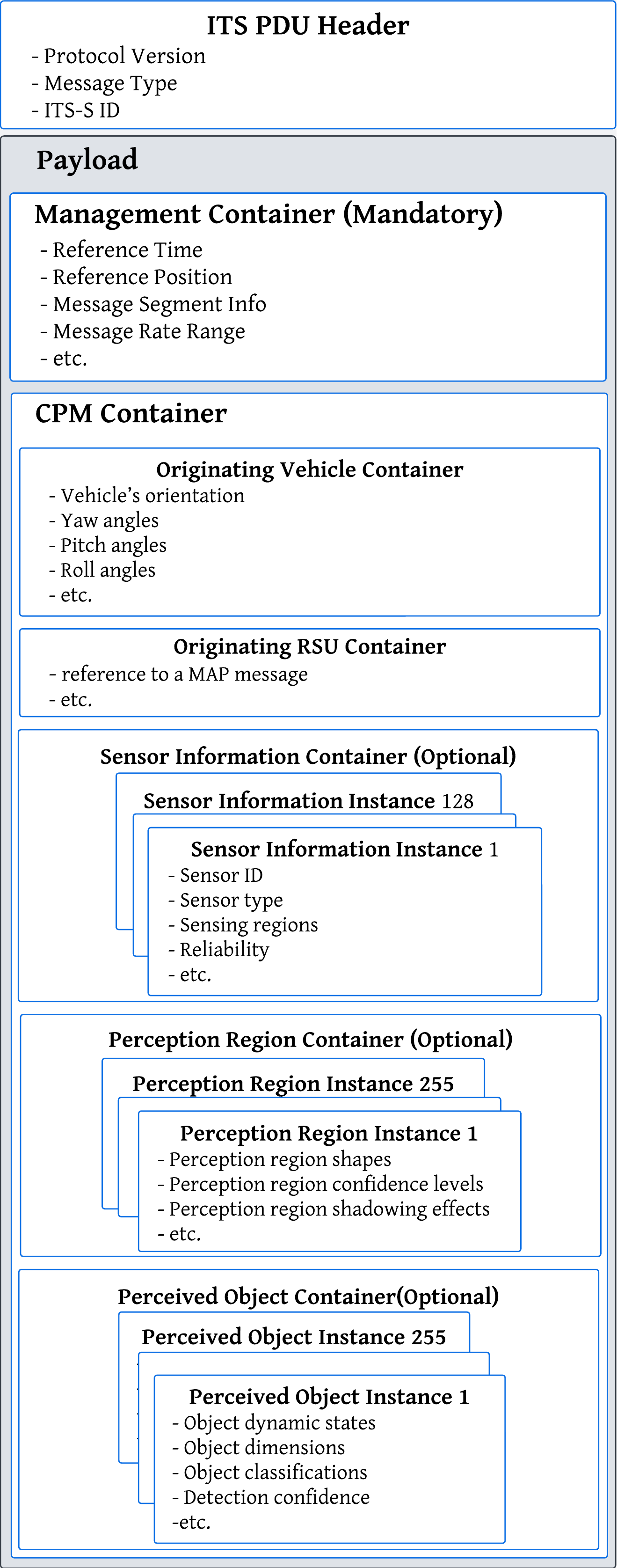}
    \caption{Definition of the ETSI CPM packet structure according to \textit{ETSI TS 103 324 V2.1.1 (2023-06)}.}
    \label{fig:CPM structure}
\end{figure}

\begin{figure*}
    \centering
    \includegraphics[width=6.5in]{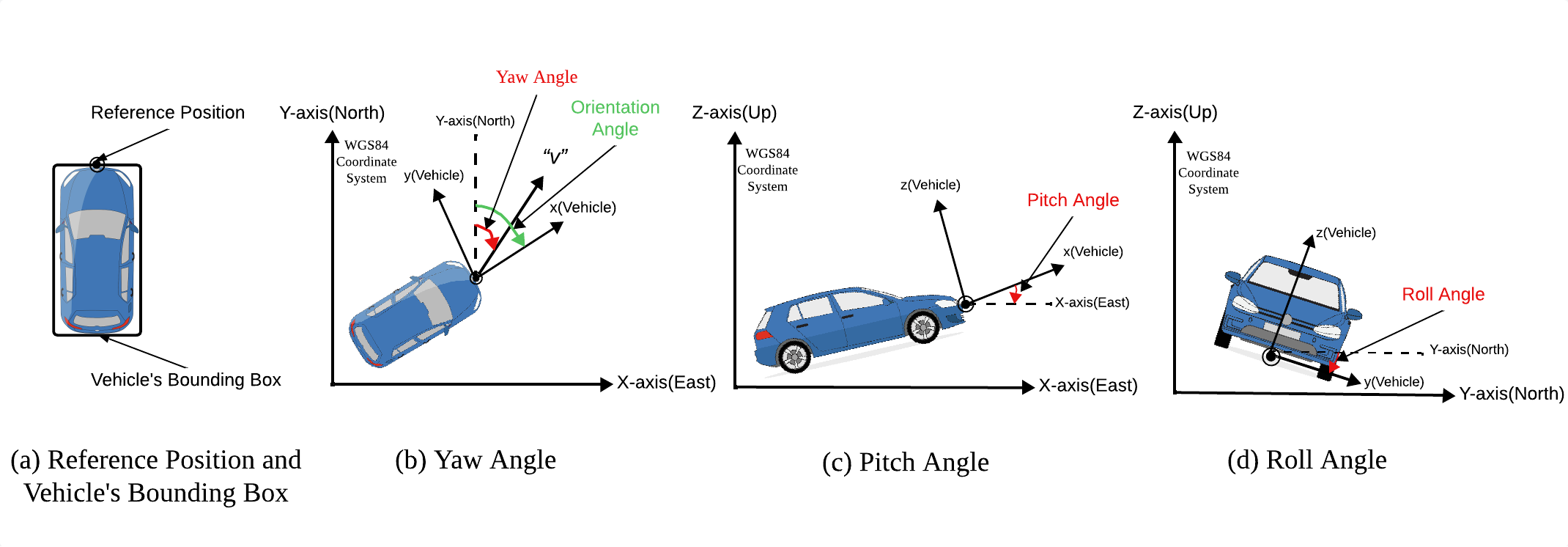}
    \caption{Visual representation of the data in (a) the CPM management container and (b-d) the originating vehicle container.}
    \label{fig:OV and PR}
\end{figure*}

\begin{figure*}[!t]
    \centering
    \includegraphics[width=6.5in]{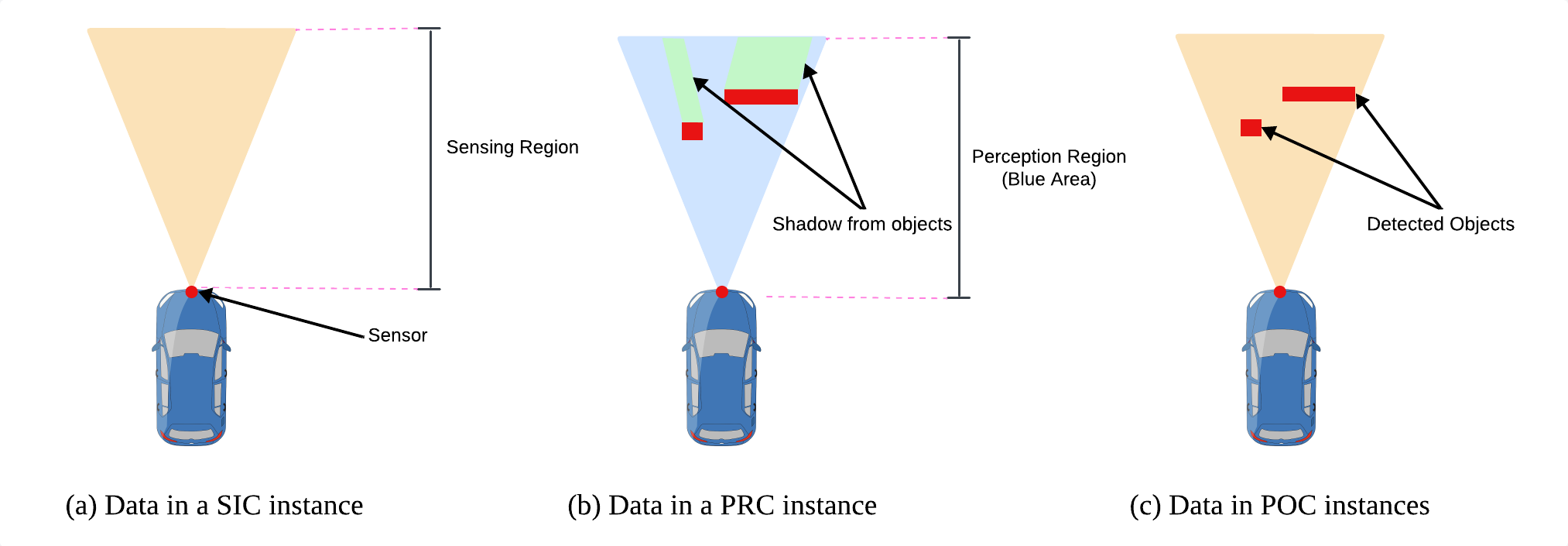}
    \caption{Visual representation of the data captured by SIC, PRC, POC instances in the CPM container.}
    \label{fig:Example_of_SIC_PRC_POC}
\end{figure*}

The most recent structure of an ETSI specified Collective Perception Message (CPM) is depicted in Fig. \ref{fig:CPM structure}. This section describes the structure of a CPM, highlighting those parts that are important in the context of redundancy mitigation. A CPM is comprised of an ITS Protocol Data Unit (PDU) header and a payload containing two constituent components: a management container and a CPM container. 
The \textbf{Management Container} is a mandatory element within the CPM. This container includes important information such as the \textit{Reference Position}, which is the ground position at the front centre of the vehicle's \textit{bounding box}, as depicted in Fig~\ref{fig:OV and PR}(a) and the \textit{Reference Time}, which is vital for synchronising timestamps within CPMs exchanged between vehicles or infrastructure. A \textbf{CPM container} consists of a number of sub-containers, some of which comprise multiple instances. An instance represents a complete set of data specific to its container type. In this paper, we will place a special emphasis on the SIC, PRC and POC sub-containers as they relate to inclusion/filtering and redundancy mitigation rules. A visual representation of the data contained in the SIC, PRC and POC is shown in Fig. \ref{fig:Example_of_SIC_PRC_POC} with the main purpose of each container now summarised.  

\begin{enumerate}
            \item \textbf{Originating Vehicle Container:} 
            Stores data related to the vehicle's orientation, including \textit{yaw} (Fig. \ref{fig:OV and PR}b), \textit{pitch} (Fig. \ref{fig:OV and PR}c), and \textit{roll} angles (Fig. \ref{fig:OV and PR}d). These represent the left/right angle and longitudinal and lateral tilt, respectively, and describe the vehicle's movement and positioning in a three-dimensional space. This orientation is measured relative to two coordinate systems: the global WGS84 system, which provides a fixed reference for real-world positioning, and the vehicle's own coordinate system. This data is crucial for accurately interpreting the reliability of the sensor information, providing essential context for the Sensor Information Container.

            \item \textbf{Originating RSU Container:}  Includes a reference to a MAP (topology) Extended Message(MAPEM) message~\cite{etsi2020103} providing contextual road infrastructure information to the CPM for enhanced situational awareness. The Originating Vehicle Container and Originating Road Side Unit (RSU) Container are mutually exclusive, depending on the transmitter type.
            
            \item \textbf{Sensor Information Container (SIC):} Stores 1-128 sensor instances. Represented in Fig. \ref{fig:Example_of_SIC_PRC_POC}(a), each instance contains data related to the sensor's specification and its sensing region (theoretical range and field of view). Such data includes its ID, type, specification, and reliability. This data is crucial for associating perceived objects with specific sensors and distinguishing idealistic sensor coverage from its actual perception.
            
            \item \textbf{Perception Region Container (PRC):} This optional container stores 0-255 perception region instances and is illustrated in Fig. \ref{fig:Example_of_SIC_PRC_POC}(b). In contrast to the idealistic sensor specification provided in the SIC, this reflects the actual area that the sensor can perceive, accounting for occlusions and including data related to shapes, confidence levels, and any shadowing effects. 
            \item \textbf{Perceived Object Container (POC):} This optional container stores up to 255 detected object instances. Each instance includes information about a single object, shown in red in Fig. \ref{fig:Example_of_SIC_PRC_POC}(c), such as its dynamic state, dimensions, classification and detection confidence.  
\end{enumerate}

Several differences exist between the CPM structure initially proposed in \textit{ETSI CPM TR (2019)} compared to the structure specified in \textit{ETSI CPM TS (2023)}. Some of the main changes include:
\begin{itemize}
    \item \textit{ETSI CPM TS (2023)} distinguishes between the 'Originating Vehicle Container' and the 'Originating RSU Container', which were collectively referred to as the 'Station Data Container' in \textit{ETSI CPM TR (2019)}.


    \item The 'Free Space Addendum Container' in \textit{ETSI CPM TR (2019)} has been replaced with a 'Perception Region Container' in \textit{ETSI CPM TS (2023)}.

    \item The 'Perceived Object Container' has been expanded to include a maximum of 255 detected objects compared to the previously defined limit of 128.
\end{itemize}

Additional changes, such as those related to Redundancy Mitigation Rules (RMR) and Value of Information (VoI) measures, are discussed in Section~\ref{RMR to VoI}.

\subsection{ETSI CPM Generation Process} 


\begin{figure}[!t]
    \centering
    \includegraphics[width=\columnwidth]{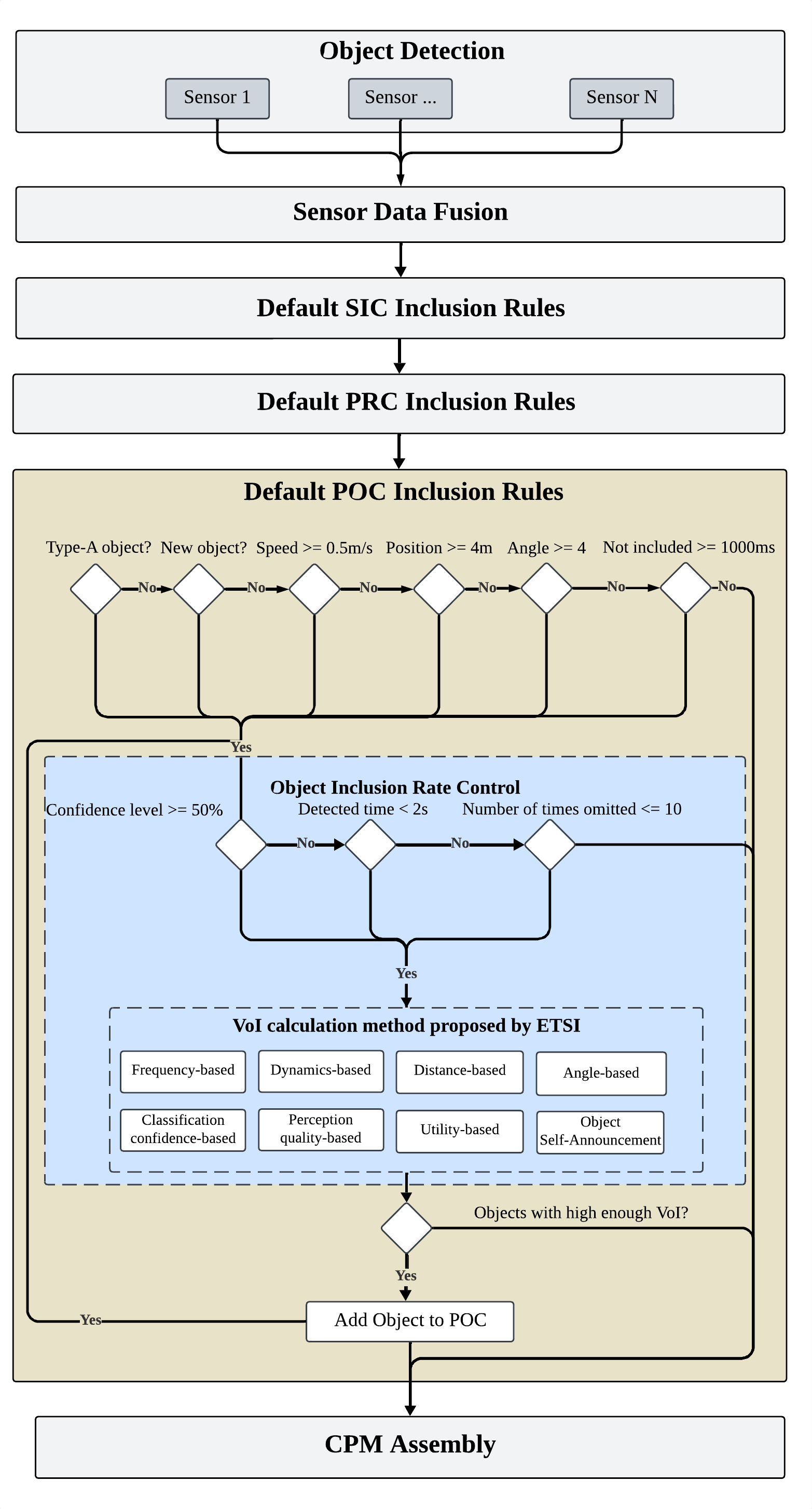}
    \caption{Summary of the CPM Generation Process including optional Object Inclusion Rules.}
    \label{fig:Flowchart}
\end{figure}

Before comprehensively reviewing the CPM inclusion rules and redundancy mitigation measures, the CPM generation process is firstly summarised as shown in Fig. \ref{fig:Flowchart} from the detection of an object to the final transmission of the CPM packet. It consists of several phases:

\textbf{Object Detection:} When a vehicle or RSU perceives an object using one or more onboard sensors, sensor fusion occurs to determine the object's state space, including its position, speed, and other relevant characteristics. The \textit{reference time} in the CPM Management Container is used to determine the delay between when the object was detected and when the CPM is assembled to ensure that the object detection age is accurately reflected in the CPM.

\textbf{Default mandatory inclusion rules:} Default inclusion rules are next applied to decide what data should be included in the SIC, PRC and POC instances of the CPM. The mandatory rules are described in Section~\ref{SIC,PRC,POC}.

\textbf{Optional Object Inclusion Rate Control:}
\label{RMR to VoI} Within the POC inclusion rules, there is an optional step that can be taken for redundancy mitigation, called object inclusion rate control. This is shown in blue in Fig. \ref{fig:Flowchart} and is based on the concept of VoI. ETSI proposes eight ways to mitigate redundancy using VoI measures, which are designed to prioritise the most relevant and important objects. A detailed discussion on this is provided in section~\ref{objRateCtrl}.

\textbf{CPM Assembly \& Transmission:} Finally the CPM is assembled and transmitted. The CPM size is constrained by the Maximum Transmission Unit (MTU) of the communication link. Thus, if the CPM size exceeds the MTU limit, the message must be segmented into smaller packets. In such cases, the SIC and PRC are included in the first CPM segment and do not need to be transmitted in subsequent segments. This approach helps reduce bandwidth usage and ensures efficient communication.
\subsection{ETSI Default Inclusion Rules}
\label{SIC,PRC,POC}
Mandatory inclusion rules are applied that govern the generation frequency of CPMs. These default mandatory rules are now summarised with the optional inclusion rate control rules described in the next section. A key concept in this process is the specification of \textit{T\_GenCPM}, which defines the time interval at which CPMs are generated. This interval is bounded by minimum and maximum thresholds, set at 100ms and 1000ms, respectively. The inclusion rules also determine which containers or instances within containers are included in each CPM packet. These rules can result in the non-transmission of CPMs, the generation of a single CPM, or the segmentation and transmission of multiple CPMs if the message size exceeds the MTU. 
\subsubsection{Default SIC Inclusion Rules}
The SIC must be included in the first generated CPM and subsequently in every CPM if the time since its last inclusion equals or exceeds 1000ms. It also must be included if an object is included in the CPM. This ensures that the sensor information remains up-to-date and provides a baseline for interpreting the other CPM data. 
\subsubsection{Default PRC Inclusion Rules}
The PRC is included in the CPM only when there are significant deviations in the dynamic perception capabilities, such as changes in perception region shape, confidence levels, or shadowing effects. These changes are compared to the static capabilities described in the SIC or to those included in past PRCs. These rules ensure that any alterations in the sensor’s perception region are communicated effectively, allowing ITS-S' to adapt to real-time environmental conditions.

\subsubsection{Default POC Inclusion Rules}
\label{Default POC Inclusion Rules}


The primary focus of this paper is on the POC and, specifically, what detected objects are included in the CPM and the frequency with which they are transmitted. The default POC inclusion rules are illustrated in Fig. \ref{fig:Flowchart} and must first classify detected objects as Type-A or Type-B. For Type-A objects, the main inclusion criterion is the frequency of detection; for Type-B objects, it is their dynamism. Default POC inclusion rules for both object types are now summarised, with no CPM generated if the inclusion rules are not met.

\noindent \textbf{Type-A Objects:} These objects typically have a higher relevance due to their potential impact on vehicular safety and navigation, e.g. a single or group of vulnerable road users (VRUs), animals, e-bikes, etc. If a new Type-A object is detected since the last CPM generation event, it is automatically included for transmission. Alternatively, if an existing Type A object has not been included in a CPM for $\geq  500ms$ (half \textit{T\_GenCPMMax}), it is included.\\
\textbf{Type-B Objects:} These objects have a lower priority and include those not classified as Type-A, e.g. other vehicles, motorcyclists, etc. If a new Type-B object has been detected, it is automatically included for transmission, otherwise it is included if it fulfills one of the following criteria:
 \begin{itemize}       
    \item  Object position change $\geq 4 m$ (absolute Euclidian distance).

     \item Object speed change $\geq 0.5 m/s$.

    \item Object heading change  $\geq 4\degree$. 
                    
     \item Time since last inclusion $\geq 1000ms$.
\end{itemize}

\subsection{Optional Object Inclusion Rate Control}
\label{objRateCtrl}

Importantly, optional POC inclusion rules can be used to further filter out included objects to reduce the size of the CPMs or the frequency of CPM generation by reducing redundant broadcasts. In \textit{ETSI CPS TR (2019)}~\cite{etsiintelligent2019}, these were referred to as \textit{redundancy mitigation techniques} based on \textit{RMRs}. In \textit{ETSI CPS TS (2023)}~\cite{etsiintelligent}, this is instead referred to as \textit{Object Inclusion Rate Control} based on \textit{VoI} measurements. The newer VoI rules employ the same techniques as specified in the older RMRs and are summarised in rows 1-6 of Table~\ref{VoI_Calculation_Methods}. However, \textit{ETSI CPS TS (2023)} added two new additional VoI measures, namely \textit{angle-based} and \textit{classification confidence-based}. 
Another key difference is that the \textit{ETSI CPS TS (2023)}~\cite{etsiintelligent} proposes combining multiple VoI measures to better manage network channel load. 
To use these optional additional rules, certain conditions must be met, as indicated in blue in Fig. \ref{fig:Flowchart}:
\begin{itemize}
    \item The confidence level of the object must be $\geq 50\%$. 
    \item The object must be detected for $\leq 2 sec$.
    \item The number of times the object has been filtered out is $\leq 10$.
\end{itemize}

\begin{table}[htbp]
\caption{Summary of the proposed ETSI VoI methods.}
\label{VoI_Calculation_Methods}
\centering
\scriptsize 
\begin{tabular}{|p{0.15\linewidth}|p{0.65\linewidth}|}
\hline
\multicolumn{1}{|c|}{\textbf{Method}} & \multicolumn{1}{c|}{\textbf{Description}} \\
\hline 

Frequency-based & The VoI of an object is determined based on how frequently other ITS-S have reported it in a given time period, i.e. lower VoI if frequently reported. Locally perceived objects are filtered out if the number of cooperatively shared CPMs related to the same object exceeds a threshold. The threshold, \textit{N\_Redundancy}, specifies the maximum number of redundant messages. 
\\
\hline
Dynamics-based & 
The VoI of an object is determined based on the distance/speed travelled by the object, with lower VoI assigned to objects with minimal changes in position or speed. Importantly, this modifies the default mandatory object position and speed change rules (based on thresholds \textit{P\_Redundancy} and \textit{S\_Redundancy}) so that instead of triggering a CPM based on a vehicle's local perception only, it also considers cooperatively shared information about the same object to account for the distance/speed travelled. This proposal is based on the Pre-Filter method proposed by Thandavaryan et al. in~\cite{thandavarayan2020redundancy} and described in Section~\ref{rmrs}.
\\
\hline
Distance-based & The VoI of an object is determined based on the Euclidean distance between the transmitting ITS-S and remote ITS-S' with whom it is cooperatively sharing CPMs about the same object in a given time period, i.e. lower VoI is assigned to objects detected by nearby ITS-S'. If the distance is below a threshold \textit{R\_Redundancy} the object is filtered out. 
\\
\hline
Object Self-Announcement & The VoI of an object is determined based on its communication capabilities and the frequency with which this object has been transmitting packets informing of its presence. Objects frequently broadcasting information about themselves are determined to have a lower VoI and may be omitted from a CPM.\\

\hline
Perception quality-based & The object VoI is based on the difference between the local object perception quality of the transmitting ITS-S and the highest object perception quality received by remote ITS-S' that have transmitted CPMs about the same object. Perception quality can be summarised as a parameter based on a combination of object age, detection confidence and detection success. Lower VoI is assigned to objects with less differences in perception quality to others.\\
\hline
Utility-based & This was referred to as entropy-based redundancy mitigation in \textit{ETSI CPS TR (2019)}~\cite{etsiintelligent2019}. The VoI of an object is based on the highest 'utility' of including the object in the CPM for all the remote ITS-S', i.e. the benefit to others of including the object. To determine such benefit, a possible utility metric is the entropy that indicates the information gained between the anticipated prior knowledge of remote ITS-S' about the locally perceived object vs their likely posterior knowledge, i.e. likely future knowledge about the object. Lower VoI is assigned to objects with anticipated low utility for remote ITS-S'. \\
\hline
Angle-based & The VoI of an object is based on the highest absolute difference between the orientation of the transmitting ITS-S to that object vs the orientation of all remote ITS-S' to the same object, i.e. lower VoI is assigned to objects that have a similar angle to the perceived object.\\
\hline
Classification confidence-based & The VoI of an object is based on the difference between the local object classification confidence at the transmitting ITS-S and the highest object classification confidence received by the remote ITS-S about the same object. Lower VoI is assigned to objects with lower differences in classification confidence between ITS-S'.\\
\hline
\end{tabular}
\end{table}

\subsection{Additional ETSI CPS Measures that utilise VoI}
\label{additionalVoIMeasures}

\textit{ETSI CPS TS (2023)}~\cite{etsiintelligent} specifies two additional uses of VoI beyond its application in the optional POC inclusion rules, i.e. Frequency and Content Management and Multi-Channel Operation (MCO).
\begin{itemize}

    \item \textbf{Frequency and Content Management:} This measure aims to use VoI to adjust the CPM generation rate based on the aggregate VoI of objects in a given region, as illustrated in Fig. \ref{fig:Inclusion rate control}. When the sum of the VoI of a group of detected objects in a given region is low, CPMs are generated less frequently but with richer information. Conversely, high VoI leads to more frequent transmission of CPMs to ensure timely updates of important information. In general, both the CPM generation interval \textit{T\_GenCPM} and the number of perceived objects and regions to be included in the CPM are chosen to maximise the overall VoI rate of the generated CPMs. The VoI rate is defined as the sum of the VoI of all objects and regions to be included in the CPM divided by \textit{T\_GenCPM}. 
    This approach allows for dynamic adjustment of CPM generation frequency, balancing the need for timely updates with network efficiency. This has yet to be quantitatively evaluated.
    \begin{figure}
        \centering
        \includegraphics[width=\columnwidth]{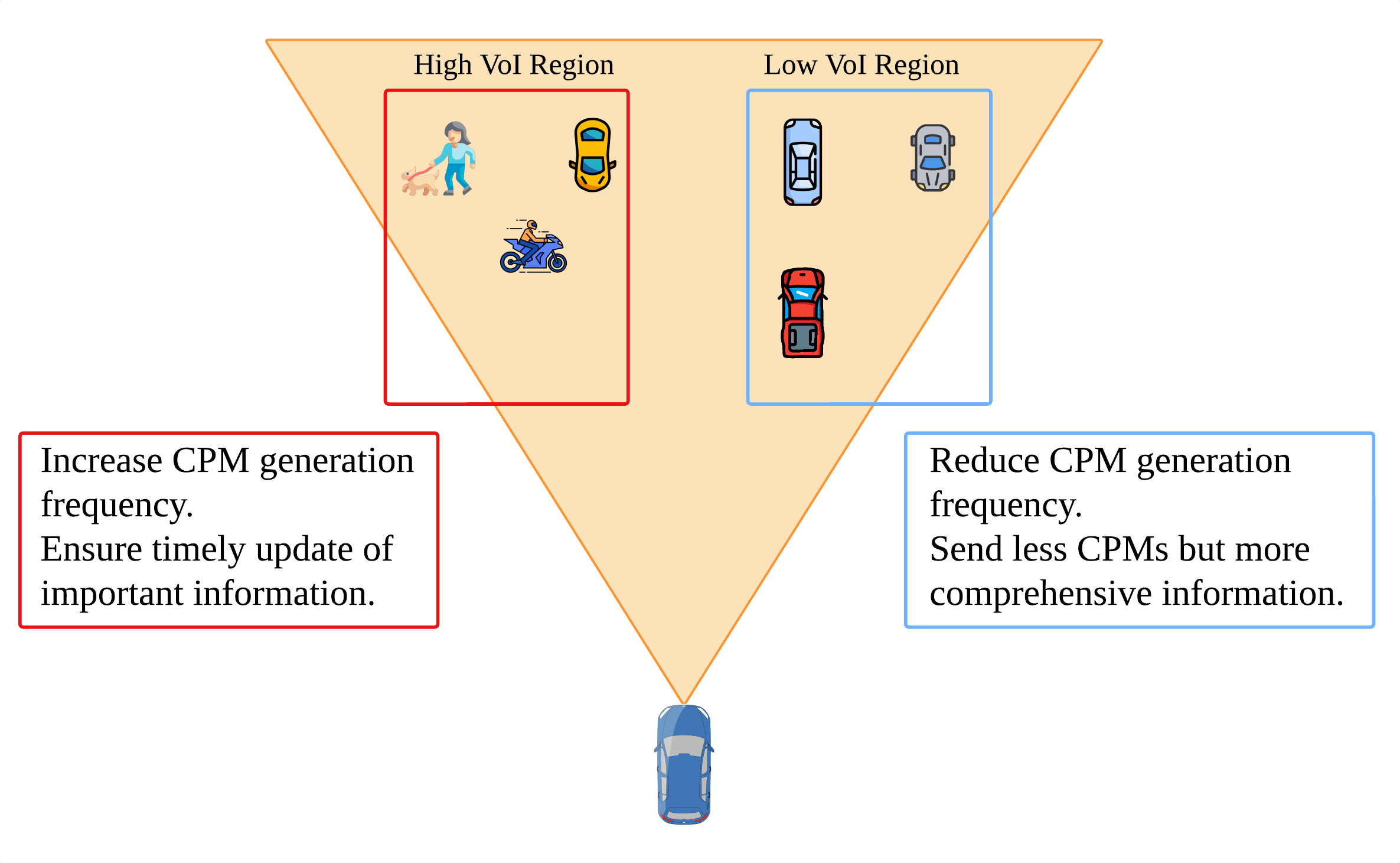}
        \caption{ETSI proposed VoI based frequency and content management.}
        \label{fig:Inclusion rate control}
    \end{figure}
    
    \item \textbf{Multi-Channel Operation (MCO):} 
    
    In \textit{ETSI CPS TS (2023)}~\cite{etsiintelligent}, MCO is proposed to manage and optimise communication channels in C-ITS environments. To contextualise, current C-ITS systems deployed in Europe follow the ETSI Release 1 system architecture~\cite{etsiintelligentC-ITS}, which accommodates 'Day 1' traffic safety and efficiency services~\cite{etsiintelligentRelase1App}. As the required data rates for such services are relatively low, a single 10 MHz radio channel was considered sufficient. However, the advent of 'Day 2' applications, e.g. collective perception along with other sensor-driven services, has motivated the need for much higher data rates and hence greater allocated spectrum as it has been shown that a single channel will not suffice~\cite{car2car2024}. This has resulted in a set of MCO standards specified across a range of C-ITS capabilities~\cite{etsiintelligent2022, etsiintelligent2022MCO,etsiintelligentMCOAccessLayer, etsiintelligentMCOStudy}. 
    Central to MCO is the \textit{Bandwidth Management Entity (BME)}, which allocates and configures radio channels based on \textit{Functional Configuration Profiles (FCPs)}, i.e. application communication requirements (preferred channels, required bandwidth, transmission range etc.) and \textit{Functional Configuration Limits (FCLs)} representing real-time network conditions (available bandwidth, transmission constraints etc). Further information about MCO can be found in~\cite{bazzi2024multi}.

    Importantly, for the collective perception service, leveraging MCO in a multi-channel environment enables the strategic utilisation of VoI to prioritise and more reliably deliver certain CPMs. Channels are dynamically allocated based on the importance attached to the entire CPM packet, i.e. the collective VoI of the packet containing multiple objects with mixed individual VoI. High VoI packets are prioritised on preferred channels, while those with lower VoI are routed to alternative channels. This method, as proposed in \textit{ETSI CPS TS (2023)}~\cite{etsiintelligent}, is proposed to ensure the efficient use of multiple channels across ITS-G5, C-V2X, and NR-V2X technologies. 

\end{itemize}

\section{Collective Perception \& the Communication Network \label{network}}


The primary motivation for adopting redundancy mitigation techniques lies in the recognition that the default CPS can potentially overload the communication network. Some researchers have aimed to quantify this issue, while others have evaluated the impact of the RMRs proposed in \textit{ETSI CPS TR (2019)}~\cite{etsiintelligent2019}. A review of the literature relating to these efforts is now provided. Notably, there have been no studies investigating the impact of the new redundancy mitigation measures introduced in \textit{ETSI CPS TS (2023)}~\cite{etsiintelligent}, specifically the angle and classification confidence based VoI methods, the VoI based frequency and content management outlined in Section~\ref{additionalVoIMeasures} or the use of MCO.

\subsection{Network Challenges posed by Collective Perception}
\label{NetworkChallengesPosedByCP}

In~\cite{thandavarayan2020redundancy} and~\cite{thandavarayan2020generation}, Thandavarayan et al. sought to quantitatively evaluate the negative impact that the CPS has on network load. In both papers, the authors considered identical scenarios (highway and urban environments), each with different traffic densities and vehicles equipped with 360° sensors.
In~\cite{thandavarayan2020redundancy}, the authors found that for a highway setting, vehicles generated an average of 9.8 CPMs per second (CPMs/s) and 9.6 CPMs/s in low and high traffic densities respectively. In the urban setting, an average of 6.1 CPMs/s and 5.7 CPMs/s were generated. They concluded that the default ETSI POC Inclusion Rules resulted in too many CPM packets being transmitted, thereby increasing the number of redundant reports of a given object and unnecessarily increasing network load. To mitigate this, they proposed a scheme called \textit{Pre-filtering}, which is detailed in Section~\ref{PF} and forms the basis for the optional ETSI \textit{dynamics based} redundancy mitigation measure. 
In~\cite{thandavarayan2020generation}, the same authors showed that 50\% to 60\% of CPMs contained 4 or fewer objects for a highway environment, while in the urban scenario, approximately 50\% of the CPMs contained only 1 object, and 90\% contained 3 or fewer objects. Thus, the evaluation concluded that the ETSI default POC inclusion rules resulted in CPMs containing too few detected objects, i.e. many small packets. The authors proposed a scheme called \textit{Look Ahead} to send fewer packets containing more objects. This is further described in Section~\ref{LA} and is included in \textit{ETSI CPS TR (2019)}~\cite{etsiintelligent2019}.




Allig et al.~\cite{allig2019dynamic}, reviewed the ETSI CPM format when it was released in \textit{ETSI CPS TR (2019)}~\cite{etsiintelligent2019}. However, the authors did not consider the specified CPM format in their quantitative evaluation, instead they simulated a similar message type to emulate a CPM based on an older proposal called the Environmental Perception Message (EPM). An EPM message mimics a CPM, except it omits the PRC container and is limited to 10 SICs and 10 POCs. The authors concluded that the CPM format is comprehensive, which ensures detailed collective perception of surrounding objects. However its complexity and size can pose significant challenges regarding network load and processing requirements. To address these issues, the authors proposed a dynamic dissemination strategy that optimises message content, described further in Section~\ref{Dynamic Dissemination}.

Recently, Xhoxhi \& Schiegg~\cite{xhoxhi2023first} analysed the spectrum requirements of CPMs under C-ITS Release 2  (\textit{ETSI CPS TS (2023)}). The authors concluded that CPMs face very high bandwidth requirements due to frequent transmissions and highly variable packet sizes, particularly when considering optional packet fields. CPM bandwidth requirements were derived as shown in Table~\ref{Bandwidth Requirements}, indicating that this can reach up to 53.48 MHz in motorway scenarios when all optional fields are used and that the channel requirements are significant for urban environments also. Given the 10 MHz allocated C-ITS spectrum within the 5.9 GHz band~\cite{etsi2018intelligent}, CPM's bandwidth requirements clearly require redundancy mitigation strategies. The authors contrast this with CAMs (requiring 3.85MHz), acknowledging that extensive research has been undertaken in the past decade proving that CAM can also overload the channel in dense conditions, and techniques have been proposed to tackle this. It is evident that the challenge posed by CPMs is significantly greater. This is mainly attributable to the optional co-variance matrix that enhances the accuracy of perceived object data. Derived from the correlation matrix, which shows normalised relationships between variables like position and speed, it reflects the scale of variation and how variables change together, offering a fuller understanding of their interactions.

\begin{table}[h]
\centering
\caption{Bandwidth Requirements for CPMs in Urban and Highway Scenarios as sourced from~\cite{xhoxhi2023first}.}
\label{Bandwidth Requirements}
\begin{tabular}{|l|p{0.35\linewidth}|l|l|}
\hline
\textbf{Scenario} & \textbf{Configuration} & \textbf{CAM (MHz)} & \textbf{CPM (MHz)} \\
\hline
\multirow{3}{*}{Urban} & Without Optional Fields & 1.49 & 5.77 \\
\cline{2-4}
& With Optional Fields & 1.71 & 19.25 \\
\cline{2-4}
& Experiential & 1.49 & 11.55 \\
\hline
\multirow{3}{*}{Highway} & Without Optional Fields & 3.85 & 11.76 \\
\cline{2-4}
& With Optional Fields & 4.49 & 53.48 \\
\cline{2-4}
& Experiential & 3.85 & 23.53 \\
\hline
\end{tabular}
\end{table}

The authors also deduce that by assuming default POC inclusion rules, CPMs are transmitted at a rate of up to 10 Hz, i.e. 10 CPMs/s. This represents a similar conclusion to the authors in~\cite{thandavarayan2020redundancy} and~\cite{thandavarayan2020generation}. Finally, they conducted a detailed study on CPM sizes and concluded that the size of the CPM can vary significantly depending on the inclusion of optional fields and the number of sensed objects. They state that this is not sufficiently reflected in academic literature with the considered CPM packet sizes in most studies typically being inaccurate e.g. in~\cite{lobo2022enhancing} where 200B is used as the median CPM size. It is stated that this value is too conservative, which under-estimates the bandwidth and capacity requirements for practical implementations. A series of equations are then proposed to derive more realistic CPM sizes that can be used in future simulation studies. These equations are based on the ETSI specification~\cite{etsiintelligent} (with and without optional fields) but also based on values suggested by Industry experts.  


\begin {comment}
\begin{table}[h]
\centering
\caption{CPM size divided by container with and without optional fields from~\cite{xhoxhi2023first}.}
\label{CPM Size}
\begin{tabular}{|m{0.3\linewidth}<{\centering}|m{0.25\linewidth}<{\centering}|m{0.25\linewidth}<{\centering}|}
\hline
\textbf{Container Name} & \textbf{Message Size (bits) without optional} & \textbf{Message Size (bits) with optional} \\
\hline
ITS-PDU & 48 & 48 \\
\hline
Management Container. & 168 & 192 \\
\hline
Originating Vehicle Container. (Opt) & 24 & 64 \\
\hline
Originating RSU Container. (Opt) & 32 & 64 \\
\hline
SIC & 104 & 144 \\
\hline
PRC & 320 & 488 \\
\hline
POC & 272 & 1968 \\
\hline
\end{tabular}
\end{table}

\begin{equation}
\label{CPM_withoutOpt}
\begin{aligned}
\text{SizeCPM\_withoutOpt [bits]} = & 240 + 104 \times \text{NrSensors} \\
& + 320 \times \text{NrPerceivedRegions} \\
& + 272 \times \text{NrPerceivedObjects}
\end{aligned}
\end{equation}

\begin{equation}
\label{CPM_withOpt}
\begin{aligned}
\text{SizeCPM\_withOpt [bits]} = & 284 + 144 \times \text{NrSensors} \\
& + 488 \times \text{NrPerceivedRegions} \\
& + 1968 \times \text{NrPerceivedObjects}
\end{aligned}
\end{equation}

From these equations, we can derive that given the minimum and maximum number on instances within the SIC, PRC and POC containers as shown in Fig. \ref{fig:CPM structure}, this results in a theoretical minimum and maximum CPM size of 43-20,564 bytes respectively without optional containers and 53-80,624 bytes respectively with optional containers. Segmentation would be required beyond 1500 bytes, representing the maximum transmission unit. Acknowledging, that it is unlikely that a vehicle will be equipped with the maximum allowable specification of 128 sensors etc, Xhoxhi et al. sought automotive experts to determine typical CPM sizes based on their experience. After extensive consultation and practical experimentation, the authors derived a simplified experiential formula for CPM packet sizes. This is shown in Equation~\ref{CPM_EXP}.

\begin{equation}
\label{CPM_EXP}
\begin{aligned}
\text{SizeCPM\_EXP [bytes]}\\ = & 575 + 52 \times \text{NrPerceivedObjects}
\end{aligned}
\end{equation}

From this formula, the size of the CPM can be between 575B and 13,835B. The authors also stated that CPMs are transmitted more frequently than CAMs and VAMs without applying redundant mitigation techniques. This higher frequency is due to the detailed and continuous updates required for environmental perception and object tracking. The authors carried out simulations in Plexe-Veins~\cite{segata2014plexe} to test the transmission frequency of the CPM, applying the \textit{default POC inclusion rules} in the CPM standard once the sensors detect either a vehicle or a VRU to determine if their current state of motion changes. The results show that the CPM transmits at a frequency close to 10 Hz (10 CPMs/s) in both urban and highway environments, this conclusion is similar to that of Thandavarayan et al. in~\cite{thandavarayan2020redundancy}.

\end {comment}

Recently, Andreani et al.~\cite{andreani2023statistical}
deduced that the size of CPMs is primarily influenced by three factors: the driving environment, the object detection accuracy of the sensors and, to a lesser extent, the message generation frequency. The authors used two real-world autonomous driving datasets that included raw data on objects sensed by an autonomous vehicle and from this, they derived a real-world model to represent CPM sizes, and transmission frequency based on the default \textit{ETSI CPS TS (2023)} CPM rules. They also proposed and evaluated a generative adversarial network (GAN) model that derives synthetic CPM sizes and frequencies based on this for two considered city scenarios. The real-world scenarios are based on the datasets NuScenes~\cite{caesar2020nuscenes} (1000 scenes from Boston and Singapore, captured using multiple sensors, including cameras, LIDAR, and RADAR for 23 object classes) and Cirrus~\cite{wang2021cirrus} (long-range 6,285 RGB-LiDAR pairs for eight object categories from both highway and urban roads). The authors deduce that the driving environment influences CPM size, with urban settings generating the largest CPMs due to the higher density of detected objects. Secondly, the authors determined that the accuracy of the object detection policy greatly impacts the CPM size. This differs from much of the literature~\cite{thandavarayan2020generation,thandavarayan2020redundancy,hakim2024stc, higuchi2019value} in this area that has sought to characterise communication challenges posed by CPMs, as they have considered idealistic object detection. The authors found that by considering more real-world object sensing, fewer objects are sensed and hence, CPM sizes are much smaller. Specifically, the authors considered \textit{'relaxed'} sensing, where an object is detected if at least one RADAR or LiDAR point is associated with it, versus \textit{'restrictive'} sensing, which requires an object to have a visibility level above 40\% and be detected by both RADAR and LiDAR sensors, or sensed by a vehicle's camera and another onboard sensor. 

Finally, the most recent related work is by Figueiredo et al.~\cite{figueiredo2024enhancing} who conducted a theoretical analysis of \textit{ETSI CPS TS (2023)} and concurred with the conclusions of Thandavarayan et al. in~\cite{thandavarayan2020redundancy,thandavarayan2020generation} that the ETSI CPS generates excessive and redundant CPMs, with each CPM containing too few objects, which in turn leads to network congestion, especially in high-density scenarios. This analysis was based on real-world experiments using data collected from the Aveiro Tech City Living Lab in Portugal.

\subsection{Evaluation of ETSI Specified Redundancy Mitigation Measures}
\label{EvaluationStandardisedRMR}
This section reviews the academic literature where researchers have sought to evaluate the effectiveness of the redundancy mitigation techniques proposed in \textit{ETSI CPS TR (2019)}~\cite{etsiintelligent2019}. New techniques proposed in academic research outside of the ETSI specification are summarised in Section~\ref{rmrs}. Delooz et al.~\cite{delooz2022analysis} investigated the performance of four of the six ETSI proposed RMRs from Table \ref{VoI_Calculation_Methods} (rows 1-4) and evaluated their impact on the performance of the vehicular network with respect to network demand (Channel Busy Ratio - CBR),  perception accuracy (Environmental Awareness Ratio - EAR) and system efficiency (Redundancy Level - RL). The aim is to minimise RL and CBR while still maintaining EAR. 
The authors firstly evaluated the performance of the \textbf{Frequency-Based RMR}, which assigns lower VoI to objects that have been frequently reported by others. Once \textit{N\_Redundancy} CPMs are received about the same object within a time window \textit{W\_Redundancy} (the authors assume $1 sec$), the ITS-S stops including the object in CPMs until the window expires and the counter resets. The authors evaluated the performance for \textit{N\_Redundancy = \{1,3,5,10,15\}}. When a low threshold is set, e.g. \textit{N\_Redundancy = 1}, the authors found that this best reduced RL and CBR while maintaining the same level of EAR. At \textit{N\_Redundancy = 15}, the effect is similar to scenarios without redundancy mitigation. Importantly, the authors identify 2 drawbacks: a) when multiple ITS-S' detect objects and share CPMs within similar short time periods, this can negate the setting of a low \textit{N\_Redundancy} parameter and b) a low \textit{N\_Redundancy} parameter paired with a high \textit{W\_Redundancy} time period may block an ITS-S from transmitting the state of a sensed object that has significantly changed over a short time period. It is our conclusion that the performance of Frequency-Based RMR is highly dependent on the tuning of key parameters. This was not considered by the authors who assumed a fixed \textit{W\_Redundancy} value.
In Fig. \ref{fig:Frequency-Based RMR}, we illustrate their dependencies. Any parameter combination marked with a red 'X' will not adequately address redundancy, with the low \textit{W\_Redundancy}, high \textit{N\_Redundancy} pairing representing the worst case, i.e. many redundant CPMs while blocking timely object updates based on the previously identified drawback. In contrast, a low \textit{N\_Redundancy}, high \textit{W\_Redundancy} combination can be useful when the object state does not change rapidly, e.g. in densely populated scenarios with many pedestrians. A combination of a low \textit{N\_Redundancy} and a low \textit{W\_Redundancy} combination may be better suited in situations where object states change rapidly, e.g. in free-flowing traffic with e-scooters, other vehicles, cyclists etc. The best way to tune and adapt these parameters to the surroundings or based on the type of perceived objects represents an open research question.

\begin{figure}
        \centering
        \includegraphics[width=\columnwidth]{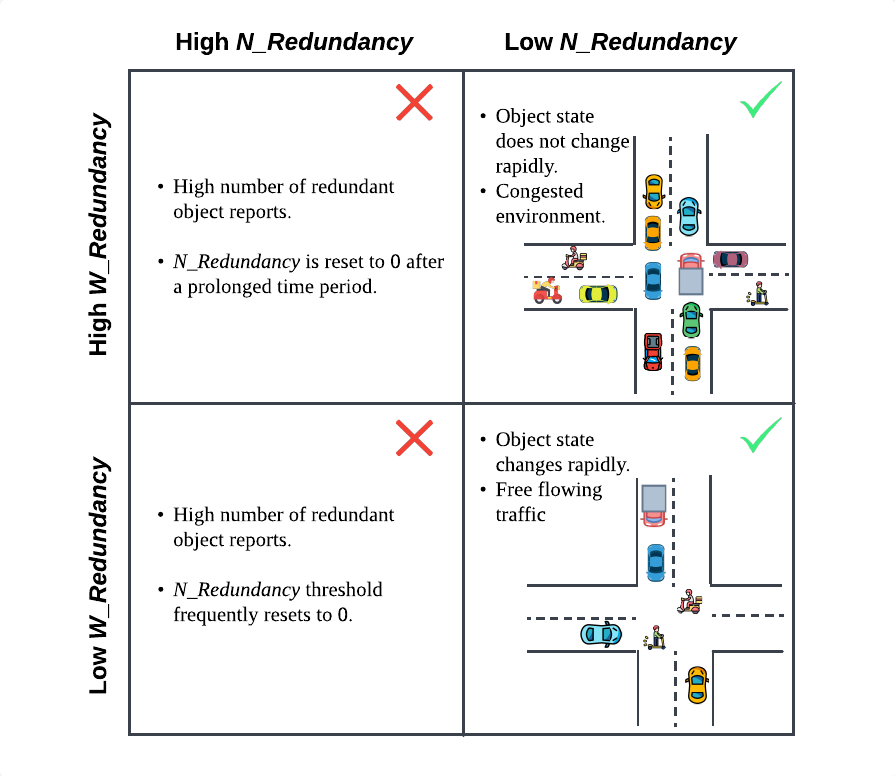}
        \caption{Impact of \textit{N\_Redundancy} and \textit{W\_Redundancy} parameters on the performance of the ETSI Frequency-Based method to effectively mitigate redundancy.}
        \label{fig:Frequency-Based RMR}
\end{figure}

\begin{figure}
        \centering
        \includegraphics[width=\columnwidth]{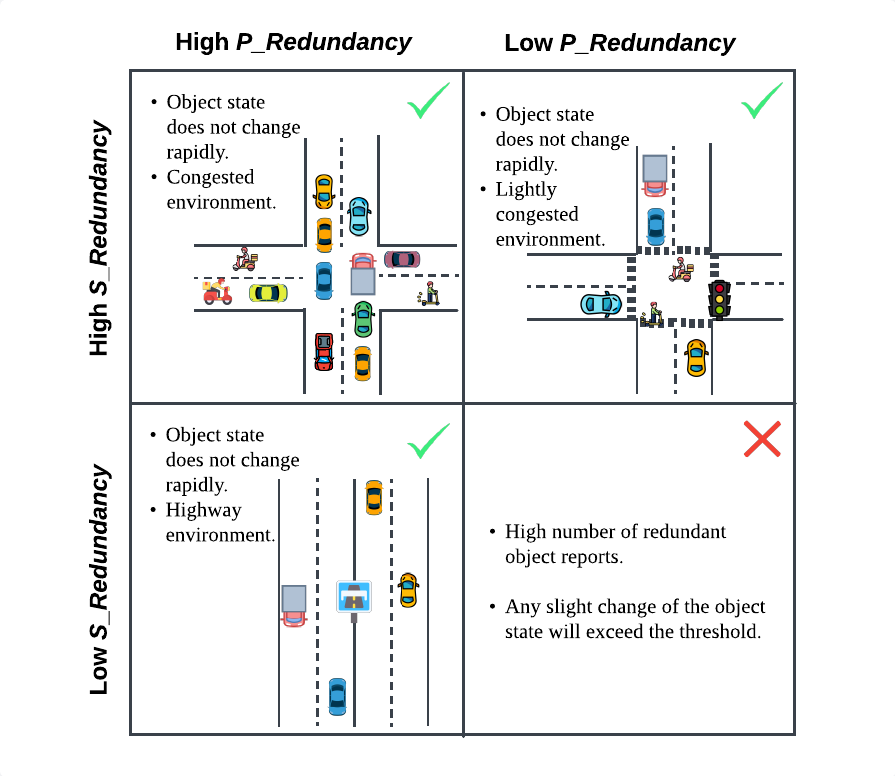}
        \caption{Impact of \textit{P\_Redundancy} and \textit{S\_Redundancy} parameters on the performance of the ETSI Dynamics-Based method to effectively mitigate redundancy.}
        \label{fig:Dynamic-Based RMR}
\end{figure}

Next, the authors evaluated \textbf{Dynamic-Based RMR}, which filters objects based on their movement. It prioritises objects with significant changes in position or velocity, i.e. assigns higher VoI to objects with large changes in position or speed. Notably, it specifically considers object data shared by other ITS-S' and not just locally perceived data, uses position and speed thresholds \textit{(P\_Redundancy} and \textit{S\_Redundancy}) and is based on the Pre-Filter method proposed by Thandavaryan et al.~\cite{thandavarayan2020redundancy}. The authors evaluated \textit{P\_Redundancy = \{2m, 4m\}} for position and \textit{S\_Redundancy = \{0.25m/s, 0.5m/s}\} for speed, during a time window of \textit{W\_Redundancy (1s)}. The authors evaluation showed a reduction in channel load of between 75-80\% while maintaining low redundancy levels. As with Frequency-Based RMR, it is our observation that the parametrisation of this scheme is paramount and is dependant on the environmental context. We illustrate this in Fig.~\ref{fig:Dynamic-Based RMR} with such parametrisation considerations not addressed in the study. Low \textit{P\_Redundancy}, low \textit{S\_Redundancy} is deemed ineffective as it generates excessive redundant updates, particularly in high-density environments. While the other three combinations can work to tackle redundancy, they work best in specific contexts. High \textit{P\_Redundancy}, high \textit{S\_Redundancy} is best suited for static/slowly changing environments where only significant changes trigger updates, minimising redundant transmissions for stationary objects. Low \textit{P\_Redundancy}, high \textit{S\_Redundancy} could be effective in urban traffic environments, where slight position changes and significant speed variations are common. This combination would ensure timely updates without overloading the network with unnecessary messages. Finally, high \textit{P\_Redundancy}, low \textit{S\_Redundancy} would work best for highway conditions, where position changes are important but speed remains relatively constant. This reduces redundant updates for fast-moving but predictable objects.

The authors next evaluate the \textbf{Distance-Based RMR}, which filters objects based on their proximity to other V2X stations. An object is excluded from a CPM if a station within the range of \textit{R\_Redundancy = \{25m, 50m, 100m, 200m\}} has already transmitted information about that object during a time window of \textit{W\_Redundancy ($1 sec$)}. This RMR significantly reduces the CBR, particularly at larger ranges like \textit{R\_Redundancy = 200m}, achieving up to 90\% reduction while maintaining a high EAR. Two limitations are observed. The performance is highly dependent on the value of \textit{R\_Redundancy}. If the range is too small, the reduction in redundancy is minimal, leading to a higher CBR. On the other hand, a large \textit{R\_Redundancy} value may result in under-reporting, particularly in complex environments where obstacles or terrain features block object detection. Secondly, vehicles located farther from the object or those with obstructed views might miss critical updates if the rule filters out transmissions from nearby stations. This can lead to situations where vehicles are unaware of objects that could be relevant for safety or navigation. The author points out that from an operational perspective, distance-based RMR is easier to implement compared to the \textit{Dynamic-Based RMR}.

Finally, the authors evaluate the \textbf{Self-Announcement RMR}, which filters objects that are capable of sending their own V2X messages (e.g., CAM), excluding them from CPMs when they self-report. This RMR is shown to be effective at higher market penetration rates. However, this also represents limitations, as it can only be effective if objects can send their own V2X messages. In mixed environments with a low proportion of V2X-capable vehicles, its effectiveness is limited, as it does not address redundancy for non-V2X objects. The authors point out that this RMR should not be considered as a standalone rule but rather as a supplement to another RMR, though how this operates in practice and in combination with other schemes remains unexplored.


Importantly, the authors conclude that all four RMRs effectively reduce channel load and have minimal negative impacts on vehicle perception but indicate that no single RMR is universally optimal; instead, each has situational strengths. \textit{ETSI CPS TS (2023)} has recommended that using multiple VoI methods together can improve overall redundancy efficiency. Very recently, two authors~\cite{sakr2024evaluation, Malik2024-bt} have evaluated such combinations. In~\cite{sakr2024evaluation}, Sakr extended the research of Delooz et al.~\cite{delooz2022analysis} to evaluate the combination of \textit{Self-Announcement-Based} and \textit{Dynamic-Based} methods as well as \textit{Frequency-Based} and \textit{Dynamic-Based} techniques. Ultimately, the authors noted that these combinations achieved better redundancy than if no mitigation technique was applied, but they achieved worse perception accuracy and channel load reduction than when considering a single mitigation scheme. It was noted, however, that the combination of schemes achieved a better Age of Information (AoI), which represents the freshness of the information available to the vehicle and is defined as the time elapsed since the last update was received about a particular object. An analysis of why the performance of the combined schemes was worse, i.e. whether they actively counteract each other, was not provided. Similarly, Malik et al. in~\cite{Malik2024-bt} also extended the study in~\cite{delooz2022analysis} by combining \textit{Self-announcement} RMR with \textit{Distance-Based} RMR. The authors made the same observation of declined performance for the combined mitigation schemes but did not analyse why this occurred. 

\subsection{CPM Performance Measurement Metrics}
After introducing the redundancy mitigation methods proposed in the ETSI standard, it is essential to clarify the key performance metrics used to evaluate CPM. As shown in Table~\ref{Measurement Metrics}, these metrics are broadly grouped into two categories, network performance and perceptual accuracy, reflecting both the communications‐related and detection‐related aspects of CPM. Because these indicators will be frequently discussed in the following sections, summarising them in advance ensures clarity and consistency when assessing the effectiveness of different redundancy mitigation techniques.

\begin{table*}[h]
\centering
\caption{Typical CPS Measurement Metrics.}
\label{Measurement Metrics}
\begin{tabular}{|l|p{0.23\linewidth}|p{0.37\linewidth}|p{0.15\linewidth}|}
\hline
\textbf{Category} & \textbf{Metric} & \textbf{Definition} & \textbf{References using this metric} \\
\hline
\multirow{7}{*}{\textbf{Network Performance}} 
& \textbf{Channel Busy Ratio (CBR)}. \textit{Also referred to as Channel Busy Time in some literature.}
& The duration the communication channel is occupied over a given time interval. 
& \cite{thandavarayan2020redundancy}, ~\cite{thandavarayan2020generation}, ~\cite{thandavarayan2023scalable}, ~\cite{chtourou2021context}, ~\cite{masuda2022feature}, ~\cite{hakim2023collision}, ~\cite{hakim2024stc}, ~\cite{lobato2023redundancy}, ~\cite{allig2019dynamic}, ~\cite{bai2023cooperverse}, ~\cite{zhou2022aicp}, ~\cite{huang2023relayed}. \\
\cline{2-4}
& \textbf{Packet Delivery Ratio (PDR)}. \textit{Also referred to as Packet Reception Ratio, Efficiency in Data Transmission and CPM Delivery Ratio in some literature.} 
& The ratio of transmitted packets that are successfully received. 
& \cite{thandavarayan2020redundancy}, ~\cite{thandavarayan2020generation}, ~\cite{thandavarayan2023scalable}, ~\cite{chtourou2021context}, ~\cite{hakim2024stc}, ~\cite{higuchi2019value}, ~\cite{bai2023cooperverse}, ~\cite{huang2020data}, ~\cite{huang2023relayed}.\\
\cline{2-4}
& \textbf{Packet Loss Ratio (PLR)} 
& The proportion of transmitted packets that are never received. 
& \cite{zhou2022aicp}, ~\cite{higuchi2019value}. \\
\cline{2-4}
& \textbf{Bandwidth Requirement} 
& The bandwidth required to maintain a service. 
& \cite{hu2022where2comm}, ~\cite{zhou2022aicp}, ~\cite{cui2022coopernaut}, ~\cite{bai2023cooperverse}, ~\cite{abdel2021v2v}. \\
\cline{2-4}
& \textbf{Size of CPM}. \textit{Also referred to as Payload in some literature.}
& The data size of each CPM. 
& \cite{thandavarayan2020redundancy}, ~\cite{thandavarayan2020generation}, ~\cite{thandavarayan2023scalable}, ~\cite{lobato2023redundancy}, ~\cite{allig2019dynamic}, ~\cite{yuan2022keypoints}, ~\cite{hakim2024stc}, ~\cite{huang2020data}. \\
\cline{2-4}
& \textbf{Transmission Delay} 
& The incurred delay between the transmitter to the receiver. 
& \cite{shen2024voi}, ~\cite{huang2023relayed}. \\
\cline{2-4}
& \textbf{CPM Generation Frequency} 
& The rate of CPM generation. 
& \cite{thandavarayan2023scalable}, ~\cite{masuda2022feature}, ~\cite{hakim2023collision}. \\
\hline
\multirow{5}{*}{\textbf{Perceptual Accuracy}}
& \textbf{Environmental Awareness Ratio (EAR)} 
& The ratio of perceived objects that an ITS-S is aware of in a given region vs the ground truth. 
& \cite{thandavarayan2020generation}, ~\cite{thandavarayan2023scalable}, ~\cite{chtourou2021context}, ~\cite{masuda2022feature}, ~\cite{hakim2023collision}, ~\cite{hakim2024stc}, ~\cite{huang2020data}, ~\cite{li2023study}. \\
\cline{2-4}
& \textbf{mean Average Precision (mAP)} 
& Evaluates both localisation accuracy (e.g. bounding box accuracy) and classification performance (e.g. object identification). This makes it crucial for measuring object detection effectiveness by calculating the mean of all average precision scores. 
& \cite{yuan2022keypoints}, ~\cite{bai2023cooperverse}, ~\cite{liu2020when2com}, ~\cite{yuan2023generating}, ~\cite{cui2022coopernaut}. \\
\cline{2-4}
& \textbf{Object Perception Rate}. \textit{Also referred to as a Cooperative Perception Accuracy (CPA) in some literature.} 
& The frequency with which objects are detected in a given time period. 
& \cite{yuan2022keypoints}, ~\cite{bai2023cooperverse}, ~\cite{liu2020when2com}, ~\cite{yuan2023generating}, ~\cite{ghnaya2023distributed}. \\
\cline{2-4}
& \textbf{Scenario Completion / Qualitative Visualisations} 
& Both measure how effectively the system performs in scenarios e.g. red light violations, either by tracking successful outcomes (Scenario Completion) or manually verifying them (Qualitative Visualisations). 
& \cite{cui2022coopernaut}, ~\cite{yuan2023generating}. \\
\cline{2-4}
& \textbf{Root Mean Square Error (RMSE)} 
& Error between predictions (e.g. position, heading) vs ground truth. 
& \cite{allig2019dynamic}. \\
\hline
\end{tabular}
\end{table*}

\section{Academic Literature for Redundancy Mitigation} \label{rmrs}

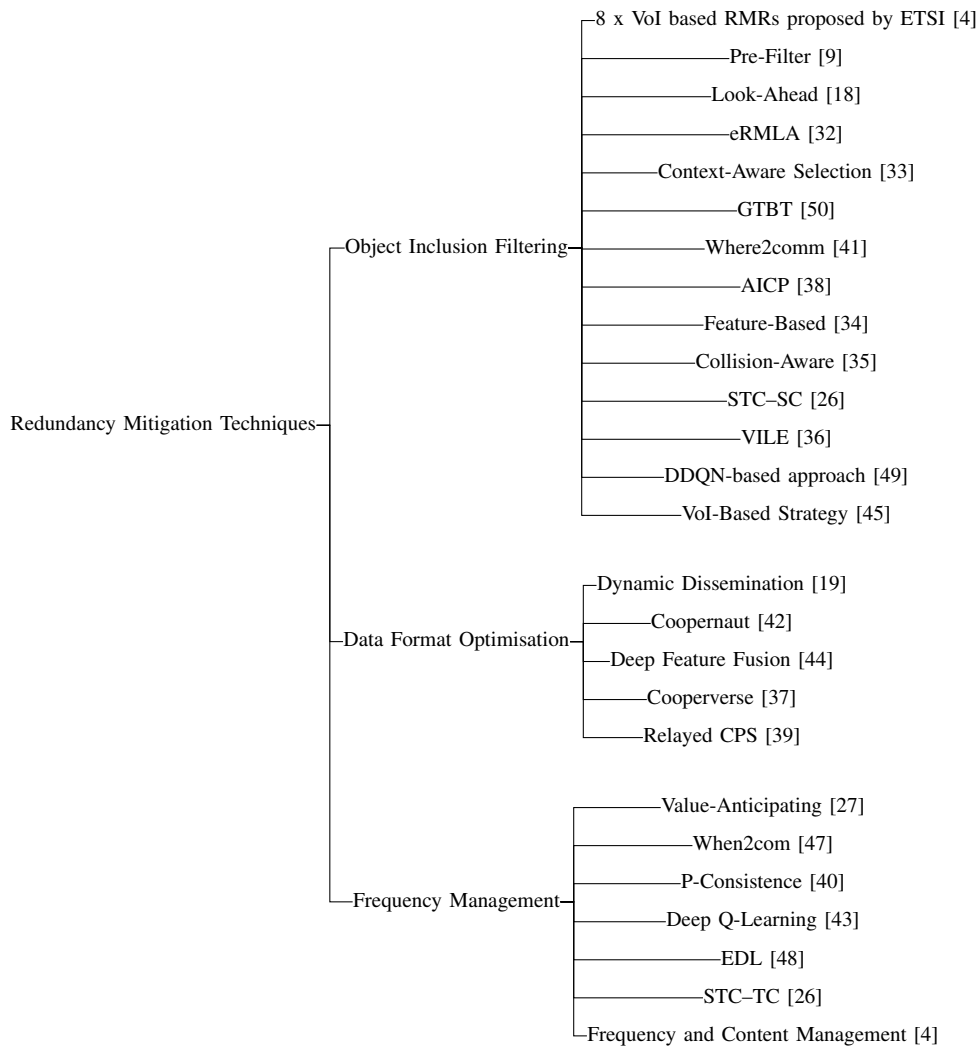
\begin{figure*}[ht]
    \centering
    \begin{forest}
    for tree={
        font=\normalfont\footnotesize, 
        grow'=0, 
        parent anchor=east,
        child anchor=west,
        edge path={
            \noexpand\path [draw, \forestoption{edge}]
            (!u.parent anchor) -- +(5pt,0) |- (.child anchor)\forestoption{edge label};
        },
        l sep=10pt, 
        s sep=5pt,  
        fit=band,
        before typesetting nodes={
            if n=1
                {insert before={[,phantom]}}
                {}
        },
        delay={where content={}{coordinate}{}},
        inner sep=0pt,
        align={center}
    }
    [Redundancy Mitigation Techniques
        [Object Inclusion Filtering
            [8 x VoI based RMRs proposed by ETSI~\cite{etsiintelligent}]
            [Pre-Filter~\cite{thandavarayan2020redundancy}]
            [Look-Ahead~\cite{thandavarayan2020generation}]
            [eRMLA~\cite{thandavarayan2023scalable}]
            [Context-Aware Selection~\cite{chtourou2021context}]
            [GTBT~\cite{chawky2022cooperative}]
            [Where2comm~\cite{hu2022where2comm}]
            [AICP~\cite{zhou2022aicp}]
            [Feature-Based~\cite{masuda2022feature}]
            [Collision-Aware~\cite{hakim2023collision}]
            [STC--SC~\cite{hakim2024stc}]
            [VILE~\cite{lobato2023redundancy}]
            [DDQN-based approach~\cite{ghnaya2023distributed}]
            [VoI-Based Strategy~\cite{shen2024voi}]
        ]
        [Data Format Optimisation
            [Dynamic Dissemination~\cite{allig2019dynamic}][Coopernaut~\cite{cui2022coopernaut}]
            [Deep Feature Fusion~\cite{yuan2022keypoints}][Cooperverse~\cite{bai2023cooperverse}]
            [Relayed CPS~\cite{huang2023relayed}]
        ]
        [Frequency Management
            [Value-Anticipating~\cite{higuchi2019value}]
            [When2com~\cite{liu2020when2com}]
            [P-Consistence~\cite{huang2020data}]
            [Deep Q-Learning~\cite{abdel2021v2v}]
            [EDL~\cite{yuan2023generating}]
            [STC--TC~\cite{hakim2024stc}]
            [Frequency and Content Management~\cite{etsiintelligent}]
        ]
    ]
    \end{forest}
    \caption{Proposed Taxonomy of Redundancy Mitigation Techniques.}
    \label{fig:rmt}
\end{figure*}

To address the network challenges posed by the CPS, researchers have proposed a variety of mitigation approaches in literature in addition to those specified by ETSI. This section critically reviews these strategies and proposes a taxonomy where redundancy mitigation schemes are classified according to \textit{object inclusion filtering}, \textit{data format optimisation}, and \textit{frequency management}. A recent survey~\cite{huang2023v2x} provided a classification based on the rule, distance and learning-based approaches. However, it is our belief that a taxonomy based on a method's conceptual approach is the preferable way to analyse the literature rather than the implementation method. The proposed taxonomic categories are now summarised, with Fig. \ref{fig:rmt} classifying the relevant literature according to the specified taxonomy. 

\begin{enumerate}
\item \textbf{Object Inclusion Filtering:} These methods filter out objects based on a variety of inclusion rules to determine relevance and reduce redundancy.
\item \textbf{Data Format Optimisation:} These methods aim to reduce the amount of information transmitted about included objects by optimising the data format within the CPM to enhance efficiency.
\item\textbf{Frequency Management:} These methods aim to send fewer CPMs by specifying novel generation and frequency management rules. 
\end{enumerate}

The redundancy mitigation measures specified by ETSI, as described in Section~\ref{additionalVoIMeasures}, also conform with this classification e.g. the \textit{VoI-based object inclusion rules} can be attributed to object inclusion filtering, and \textit{frequency and content management} can be categorised as frequency management within the proposed taxonomy. The exception to this is Multi-Channel Operation (MCO), which does not strictly fit the classification as it does not address redundancy directly but rather prioritises particular network resources as a way to negate the impact of channel congestion by preferentially transmitting high VoI packets on a preferred, least congested channel. Each scheme within the proposed taxonomy is now described.

\subsection{Object Inclusion Filtering Approaches}
\label{Literature review}
This category of redundancy mitigation techniques aims to reduce the number of objects included in each CPM, as depicted in the bottom-left portion of Fig. \ref{fig:Illustration_of_RM_Taxonomy}. Several studies have proposed methods to selectively include only the most relevant or critical objects in the CPM. 

\subsubsection{Pre-filter (PF) Method}
\label{PF}
\begin{figure}
    \centering
    \includegraphics[width=\columnwidth]{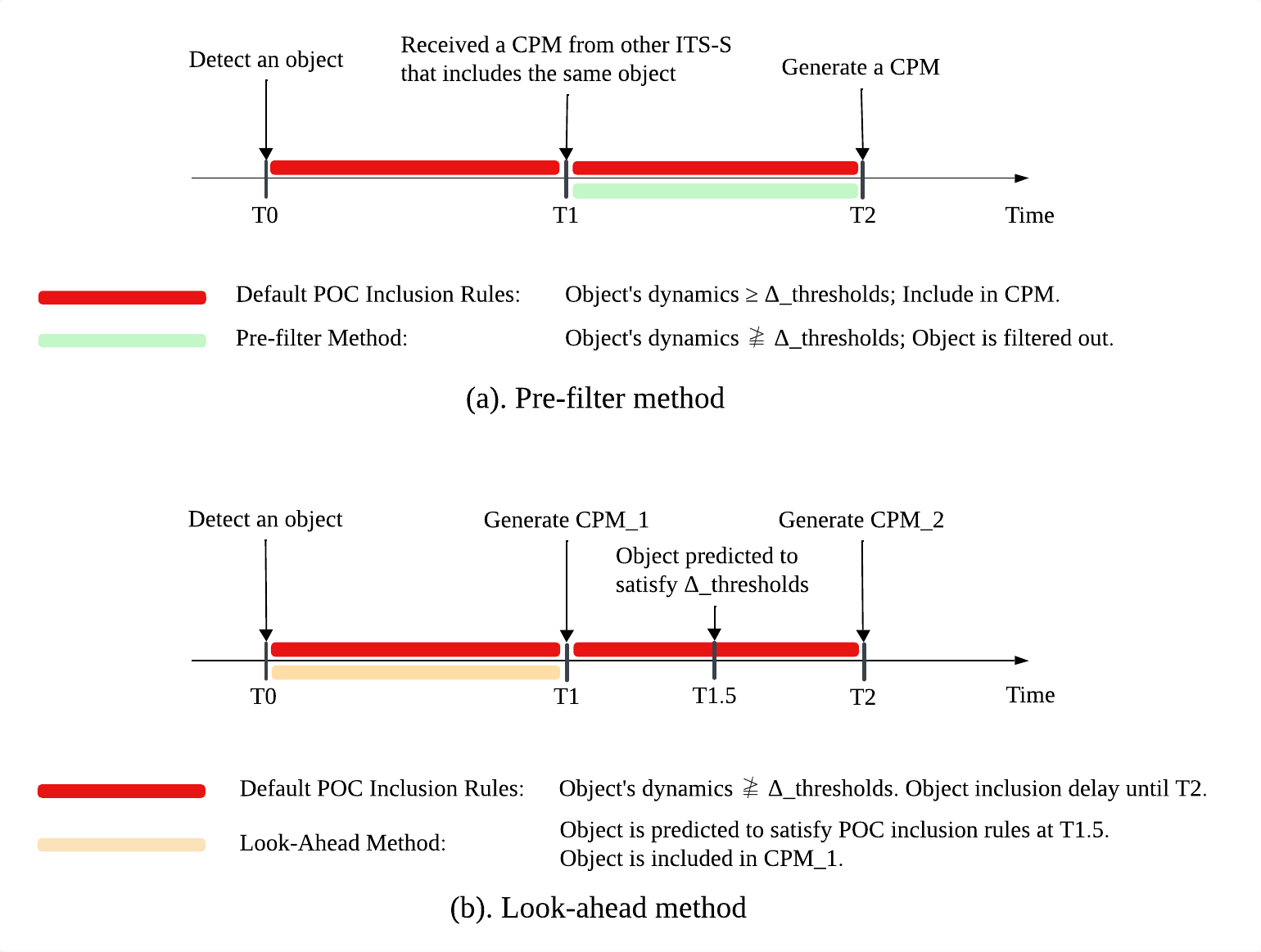}
    \caption{Schematic of Pre-filter~\cite{thandavarayan2020redundancy} and Look-ahead~\cite{thandavarayan2020generation} methods when compared to default POC inclusion rules.}
    \label{fig:Pre-filter&Look-Ahead}
\end{figure}

The \textit{PF} method, proposed by Thandavarayan et al. in May 2020~\cite{thandavarayan2020redundancy}, aims to reduce the number of redundant objects included in a CPM, with this proposal forming the basis of the \textit{"dynamics based"} VoI method specified in \textit{ETSI CPS TR (2019)}~\cite{etsiintelligent2019}. It is illustrated in Fig. \ref{fig:Pre-filter&Look-Ahead}(a).
\textit{PF} modifies the default mandatory object position and speed change rules so that instead of triggering a CPM based on a vehicle's local perception only, it also considers cooperatively shared information about the same object. The default rules compare the change in distance/speed of the same object in the interval between the previous and current CPM to determine if it exceeded a threshold. In contrast, \textit{PF} will account for the distance/speed travelled by the object since it last received an update on its position i.e. from a CPM shared by a neighbouring vehicle. The goal is to prevent the redundant inclusion of objects already reported by other vehicles, as shown in Fig. \ref{fig:Pre-filter&Look-Ahead}(a). This proposal was included in \textit{ETSI CPS TR (2019)}~\cite{etsi2019intelligent} as a potential solution with the quantitative evaluation indicating that the CBR decreases notably with a reduction between 10.1\% - 31.8\%, depending on the threshold setting.

\subsubsection{Look-Ahead (LA) Method}
\label{LA}
The same authors also introduced the \textit{LA} method~\cite{thandavarayan2020generation} in December 2020. The authors use simulation to identify an issue with the default ETSI inclusion rules where most CPMs include very few objects, leading to network inefficiencies. The proposed \textit{LA} method addresses this by modifying the default POC inclusion rules to forecast an object's future dynamics and relevance, using their known position, speed, and acceleration. The goal is to ensure that only potentially important objects are included in generated CPMs but also to reduce the number of CPMs transmitted by increasing the number of objects reported in a given packet, as shown in Fig. \ref{fig:Pre-filter&Look-Ahead}(b). The authors' evaluation indicated a maximum CBR reduction from 49.4\% to 41.4\% by increasing the number of objects in each CPM up to a maximum of 109.8\%. \textit{LA} was also included in \textit{ETSI CPS TR (2019)}~\cite{etsi2019intelligent} based on the authors findings.

\subsubsection{Enhanced Redundancy Mitigation Look-Ahead (eRMLA) Method}
\label{eRMLA}
Both \textit{Pre-filter}~\cite{thandavarayan2020redundancy} and \textit{Look-Ahead}~\cite{thandavarayan2020generation} have demonstrated benefits for CPM redundancy mitigation when individually applied. \textit{PF} reduces the number of CPMs transmitted in each second by filtering out redundant objects that have been reported by others. In contrast, \textit{LA} reduces the frequency of transmission by including more objects in each CPM. However, these schemes can work against each other when applied simultaneously. In their recent work ~\cite{thandavarayan2023scalable}, Thandavarayan et al. sought to quantify this impact and also proposed a scheme in which they could work in harmony. They evaluated two combinations: RMLA (\textit{PF} occurs first, followed by \textit{LA}) and LARM (\textit{LA} first followed by \textit{PF}). In RMLA,  \textit{PF} could remove the object if it has been cooperatively shared by another vehicle, however \textit{LA} could add the redundant object back into the CPM if it predicts it will have met the movement threshold before the next CPM is issued. In LARM, the opposite occurs, which can result in frequently transmitted CPMs with fewer objects. 

The authors, therefore, propose \textit{eRMLA}, which attempts to trade-off between reducing the frequency with which CPMs are sent and ensuring that when they are sent, there is maximum object inclusion to maintain comprehensive perception. eRMLA first applies the default generation rules and then removes redundant objects using \textit{PF}. If no object remains, the CPM is not generated. If at least one object is not filtered out by RM, the \textit{LA} mechanism re-evaluates all objects, including those initially filtered out by \textit{PF}. 
By avoiding the drawbacks of \textit{RMLA} and \textit{LARM}, \textit{eRMLA} outperforms the baseline approach (no redundancy mitigation) in four evaluation aspects: improved object perception (10\%-42\%), decreased CBR (50\%), increasing the average number of objects included in each CPM (170\%), and decreasing the number of CPMs generated per second (76\%) across all traffic densities. 



\subsubsection{Context-Aware Selection Method}
\label{Context-Aware Selection}
Chtourou et al.~\cite{chtourou2021context} propose an approach called \textit{context-aware selection} that proposes three algorithms to filter objects based on the network load and the existence of roadside infrastructure. The first algorithm, entitled\textit{CBR-Binary}, filters out all objects when the CBR is above a certain threshold, i.e. no CPMs are sent. Secondly, the \textit{CBR-Selective} scheme proposes a variation on the ETSI frequency-based RMR mechanism summarised in Table~\ref{VoI_Calculation_Methods}. However, instead of filtering out objects when they have been cooperatively shared by others beyond a static threshold \textit{N\_Redundancy}, this scheme instead specifies an adaptive \textit{N\_Redundancy} threshold based on the CBR. If the CBR is high, \textit{N\_Redundancy} is lowered to filter out objects with lower VoI. Finally, the \textit{CBR \& Infra-Selective Scheme} extends the \textit{CBR-Selective} scheme to also exclude objects announced by roadside infrastructure. 

The authors evaluation, for a 100\% penetration rate, showed that the \textit{CBR-Binary} and \textit{CBR-Selective} schemes maintained a median PDR of {$\sim$}{75\%} versus a median PDR of {$\sim$}{50\%} if no mitigation measures were applied. This is mainly attributable to reducing the CBR from {$\sim$}{80\%} to {$\sim$}{60\%}. Notably, comparable environmental awareness was maintained. There are negligible differences in performance between the two schemes or performance benefits shown at lower penetration rates. The \textit{CBR \& Infra-Selective} scheme achieved close to 100\% PDR across both urban and highway scenarios while improving the median CBR by 30\% in the presence of roadside infrastructure, assuming a 100\% penetration rate.

\subsubsection{Game Theory Based Transmission (GTBT) Method}
\label{GTBT}
In 2022, Chawky et al. proposed the GTBT method~\cite{chawky2022cooperative} to mitigate redundant message content in fog-based vehicular networks, leveraging fog nodes at the network edge for low-latency processing and efficient data management. The goal is to reduce duplicate messages and ensure that only the most useful and unique information is shared between autonomous vehicles (AVs). The authors do this by proposing a game theoretic approach (\textit{GTBT}) to decide which perceived objects should be transmitted by AVs. They define a non-cooperative game, where each AV acts as an independent player making local decisions to independently evaluate the importance of the information that it has detected (locally and via cooperatively shared information) based on factors such as potential physical collision risk and whether the object is already visible to other AVs through previously received CPMs. Potential collision risk is determined by calculating the time until a potential physical collision with other vehicles based on their current trajectories and speeds. 
To assess whether other AVs have seen the object, each AV estimates the visibility of the object using its sensors and shared information from previous messages. The importance of an object is then scored based on its relevance and urgency, considering these factors. By calculating scores for each piece of information and constructing payoff matrices, AVs can make strategic decisions to either include or not include certain objects in their messages. Essentially, the \textit{GTBT} method ensures that if multiple AVs detect the same object, they coordinate implicitly (without direct communication) by independently evaluating who should send the information based on their scores. This implicit coordination happens through the scoring and decision-making process, which helps to avoid duplication.

In their quantitative evaluation, the authors show that \textit{GTBT} reduces redundancy and improves the transmission balance of key objects. \textit{GTBT} is compared with three other methods: \textit{Max Score Based Transmission (MSBT)} - the decision to include an object is based on the highest score only i.e. its importance, regardless of whether other vehicles are potentially also cooperatively sharing this object. \textit{Random Transmission (Rand)} - randomly decides to transmit the CPM or not, leading to inconsistent results. \textit{Geo-filtering Transmission (Geo)} - objects are included based on geographic location, i.e. distance to the transmitter, which often results in the omission of important information. \textit{GTBT} reduces duplicate information by 10.5\% compared to \textit{MSBT}, provides more consistent and valuable perception dissemination compared to \textit{Rand}, with a higher value of unique information sent, and sends only 5.3\% of missed maximum score messages compared to 15.2\% for \textit{Geo}, while maintaining unique message delivery.

\subsubsection{Where2comm Method}
Hu et al. propose an approach~\cite{hu2022where2comm} based on an entirely different collective perception framework that deviates from the ETSI specified approach. While it aims to reduce the amount of raw sensor data transmitted, its main purpose is to improve collective perceptual accuracy. It proposes a means of identifying perceptual information that is spatially sparse yet important for neighbouring vehicles by devising a spatial confidence map based on a vehicle's own sensor readings and information shared by others. It then cooperatively shares a \textit{spatially sparse feature map} which contains a \textit{request map} for regions that it needs to know more about from others and a \textit{feature map} for objects that it is confident about and believes are of importance to receivers. 
The performance evaluation focused on perception accuracy, although it alludes to enhanced communication efficiency when compared to \textit{When2com}, a method proposed by different authors and described in Section~\ref{When2com}. The authors of this paper have classified \textit{When2com} as a \textit{frequency management} approach. \textit{Where2comm} outperformed previous methods, showing significant reductions in the message sizes transmitted between agents, expressed as a log scale. Specifically, results across multiple datasets have shown a 25.81\% increase in Average Precision (AP) while reducing communication bandwidth usage.


\subsubsection{Augmented Informative Cooperative Perception (AICP) Method}
Zhou et al. propose a method entitled \textit{AICP}~\cite{zhou2022aicp} that includes two aspects; a redundancy mitigation measure, acting as a transmitter redundancy filter, and a technique to filter out irrelevant objects at the receiver side, functioning as a receiver relevance filter. \textit{AICP} classifies sensed objects as 'near distance,' 'medium distance,' or 'far distance.' Only 'near distance' and 'medium distance' objects are included in a transmitted CPM to reduce its size. This approach is similar to the ETSI \textit{'distance-based'} RMR but uses two distance thresholds instead of one. The main novelty of the paper, however, comes at the receiver side. Mahalanobis distance is used to calculate the relevance of the received objects to the driver. This distance calculation measures the correlation among features and accounts for their variances and co-variances, ensuring that only the most relevant objects are displayed on a driver's augmented reality display to prevent driver information overwhelm. The \textit{Contextual Multihop Routing (CMR)} protocol within AICP initially broadcasts the collaborative message to all surrounding vehicles. However, it restricts the information from being processed and forwarded only by vehicles travelling in the same direction as the ego vehicle. Using the proposed techniques, the AICP method effectively reduces the number of packets transmitted and minimises channel load, e.g. reducing channel busy time by 81\% compared to hop limitation routing only.




\subsubsection{Feature-Based Method}
Masuda et al. ~\cite{masuda2022feature} proposed a \textit{feature-based} method that modifies the existing CPM default POC Inclusion Rules. It employs a \textit{Vehicle Identification Framework} that leverages the visible features of vehicles captured by onboard sensors e.g. colour, brand, and plate number, and adds these into the Station ID field of the CAM or CPM. At the receiver side, a process of \textit{Feature Extraction and Data Fusion} is conducted. This is where the supplemental CPM visual data is compared with information stored in the vehicle's Local Dynamic Map (LDM), ensuring accurate identification of each vehicle by fusing V2X data with locally sensed data. The LDM is a local database that maintains both local and cooperatively shared data from nearby vehicles. To assist in CPM redundancy, the system filters out objects by excluding vehicles already identified in the LDM from inclusion in future CPMs. This filtering mechanism ensures that only new or non-identified objects are included in the CPM. The performance evaluation compared against default ETSI POC inclusion rules. The \textit{feature-based} method demonstrates a superior trade-off between network load and environmental awareness, particularly at higher penetration rates of 80\% and above, achieving a 72\% reduction in CBR compared to the default ETSI method. This is because, at higher penetration rates, more vehicles are sensor equipped and can identify the unique features of neighbouring vehicles and hence filter these from their CPMs. However, at low and medium penetration rates, the default ETSI POC inclusion rules perform better. 


\subsubsection{Collision-Aware Method}
Hakim et al. proposed a clustering algorithm~\cite{hakim2023collision}  that considers collision avoidance as a key metric in reducing message redundancy while avoiding accidents. A scoring function is used to prioritise messages, ensuring that important objects are shared while minimising communication overhead. Four different scoring functions are explored to evaluate their effectiveness in clustering perceptual objects: \textit{Direction difference (S1)}, which measures the difference in heading between the sender and the perceived object; \textit{Close distance (S2)}, which prioritises nearby objects to ensure they remain in the same cluster; \textit{Far distance (S3)}, which extends the perception range by considering distant objects; and \textit{Potential collision time (S4)}, which considers the time of potential collision between the receiver and the sensed object. Each CAV performs this algorithm every 0.1 seconds and scores each perceived object according to the chosen scoring function. Objects uniquely perceived by a CAV are included in its cluster, and for objects perceived by more than one CAV, only the highest scoring transmitter continues to send information about it in subsequent intervals, thus reducing redundancy. With the exception of \textit{S2} and \textit{S3}, which are mutually exclusive, these scoring functions can be used in combination, such as \textit{S1+S2+S4} or \textit{S1+S3+S4} by assigning weights to them separately and then normalising them. Such combinations were also evaluated.

The authors quantitatively compared the performance of the proposed \textit{Collision-Aware} clustering algorithm to the existing \textit{LA} and \textit{PF} methods described earlier in this section. The evaluation showed a significant improvement in communication effectiveness. The \textit{S4} scoring function achieved the highest increase in new information, with a 603\% improvement in urban scenarios and a 509\% improvement in highway scenarios relative to the ETSI default POC inclusion rules. Additionally, the algorithm reduced the number of transmitted packets by 22.2\% in urban areas and 23.9\% on highways while increasing the number of successfully received packets by 10.6\% and 25.7\% respectively. Combined scoring functions such as \textit{S1+S2+S4} and \textit{S1+S3+S4} also showed significant improvements, particularly in highway scenarios, with the latter achieving a 44\% increase in new information. However, in urban scenarios, the combined method (\textit{S1+S2+S4} and \textit{S1+S3+S4}) show a decrease in new information (-13\% and -9\% respectively), suggesting that the individual benefits of S4 may be diluted when combined with other scoring functions in more complex environments.

\subsubsection{STC Method -- SC}
\label{SC}
The same authors who developed the \textit{Collision-Aware} method have recently further optimised it by introducing the \textit{STC} method~\cite{hakim2024stc}, which includes \textit{Spatial Clustering (SC)} and \textit{Temporal Clustering (TC)}. \textit{SC} belongs to the Object Inclusion Filtering classification, while \textit{TC} belongs to the Frequency Management classification, and the \textit{TC} will be presented in Section~\ref{TC}. The SC method specifically addresses the clustering of objects based on their spatial distribution to reduce redundant data transmission. SC uses a similar approach to GTBT by calculating the importance of objects to each CAV through parameters such as position, velocity, and potential collision risk. Objects are grouped based on their spatial proximity using a clustering algorithm that evaluates the distance between objects and their relevance to the vehicle's immediate environment. Each CAV calculates an importance score for each detected object, considering factors like time to potential collision and visibility probability. Depending on the score, different objects will be assigned to different CAVs, and the CAVs will only include the objects assigned to them in the POC.

In evaluating the \textit{SC} method, the authors found improvements in communication efficiency. The \textit{SC} method decreased communication payload by approximately 20\% and increased information reception over an infrastructure-less V2V network by 10\% compared to the ETSI standards. When combined with the \textit{LA} method, SC demonstrated further improvements. The integration of \textit{SC with LA (SC+LA)} reduced the payload by an additional 31.4\% compared to using LA alone. Similarly, the integration of \textit{SC with the RM method (SC+RM)} resulted in benefits. The SC+RM approach reduced the payload by 30\%, which is a substantial improvement over RM alone. The SC+RM method also showed a 19\% to 46.4\% increase in the number of locally perceived objects reported, depending on the scenario. Moreover, this combination enhanced the object perception ratio, showing an increase of up to 36.3\% in the reception of new, previously unperceived objects per second relative to the ETSI default inclusion rules.

\subsubsection{VILE}
Lobato et al. propose a region-based object inclusion filtering method~\cite{lobato2023redundancy}. The method works by dividing the environment into hexagonal regions and selecting a single vehicle within each region to broadcast perception data. This selection is based on proximity to the region’s centre. Vehicles periodically share information like position, speed, and direction, enabling an RSU to build a dynamic map of vehicles and select the most suitable representative for each region. This approach ensures that only unique and relevant data is shared, minimising redundancy. The authors compare VILE to baseline ETSI CPM generation rules using the \textit{Veins} framework. The simulations were conducted in a 1km² Manhattan grid scenario with vehicle densities of between 100 and 200 vehicles per km², a vehicle Field of View (FoV) of 360° and a sensor perception range of 150m. Results demonstrated that VILE improved network efficiency, reducing the number of CPMs generated by up to 55\% and redundant objects included in CPMs by up to 75\%, compared to the baseline ETSI method. 

\subsubsection{DDQN-based approach}
Ghnaya et al., in~\cite{ghnaya2023distributed} propose a Distributed Double Deep Q-Learning (DDQN) approach that aims to optimise CPM content selection in a distributed multi-CAV environment. It addresses the issue of redundant information exchange by formulating the usefulness of perceived objects as a mathematical maximisation problem, taking into account various perception contexts, including object distance, viewing angle, size, and occlusions caused by other road users. Each CAV acts as an independent agent that learns the optimal policy for selecting which perceived objects to include in the CPMs, ensuring that only the most relevant information is shared with neighbouring vehicles. The learning process utilises two deep neural networks, a training network and a target network, to achieve stable convergence and minimise overestimation bias during the training phase. The incorporation of an experience replay mechanism further enhances learning stability by allowing each CAV to learn from a diverse set of historical experiences.

The authors compare against three existing methods: \textit{ETSI default CPM Generation Rules}, which specify when and how to generate CPMs based on pre‐defined triggers; a \textit{Dynamics‐Based Redundancy Mitigation} technique, which dynamically adjusts CPM content based on changes in object positions and speeds; and a \textit{CBR‐Selective Scheme} (also referred to as \textit{Context‐Aware Selection} in~\ref{Context-Aware Selection}), which uses channel busy rate thresholds to manage CPM transmission. Simulations conducted using Artery~\cite{riebl2015artery} and SUMO~\cite{krajzewicz2010traffic} model communication and mobility across a 10km\textsuperscript{2} area of Bordeaux, France, featuring both urban and highway scenarios. Each CAV is equipped with GPS/GNSS, 360° radar, and lidar sensors with a 100m sensing range, exchanging CAMs every 0.1s and CPMs every 0.15s via an ITS‐G5 enabled 500~m coverage area.

DDQN reduces object redundancy at shorter distances (up to 150m) by approximately 10\% compared to the Dynamics‐Based approach and 15\% compared to the CBR‐Selective Scheme. It improves object perception by approx. 10\% and 15\% over the Dynamics‐Based and CBR‐Selective methods, respectively, at distances below 100m. This enhancement is crucial for safety in short‐range communication scenarios, where accurate situational awareness is especially important. Finally, the proposed method achieves a higher PDR at distances below 200~m, improving reliability by 5\% to 10\% compared to baseline techniques. 

\subsubsection{VoI-Based Strategy}
Shen et al. propose a method~\cite{shen2024voi} to address redundant CPMs when multiple CAVs approaching a road intersection within base station (BS) coverage perceive the same object. The key idea is to optimise which perception information a CAV should share with the BS to minimise redundancy. The authors define the VoI as the value of perceived information calculated by each CAV using a set of parameters (AoI, distance, FoV, and potential obstacles). Initially, each CAV sends a request containing the calculated VoI to the BS. Based on the received requests, the BS employs a simulated annealing (SA) algorithm to explore the solution space and gradually reduce the probability of accepting sub-optimal solutions during the iteration process to determine the optimal solution. The optimisation problem, modelled as a maximisation problem, aims to maximise CPS utility while minimising redundancy. Once the BS has identified which CAV needs to share the CPM, it communicates the decision to the corresponding CAV, which then shares the CPM as directed.

The authors use SimuLTE (an extension of OMNet++) and SUMO to evaluate their approach and compare it with three other methods; \textit{ETSI CPS 2019}, in which all objects are included in the CPM; \textit{random}, in which objects are included in the CPM at random; and \textit{only consider VoI}, in which no algorithm is used, and only the magnitude of the VoI value is considered. The authors show that the proposed VoI-Based strategy reduced the average transmission delay by 22.3\% compared to \textit{ETSI CPS 2019} and improved perception quality by 21.6\% compared to \textit{only considering VoI}. At the same time, all perform better than the \textit{random strategy}. The authors also compared their approach with Dung Beetle Optimisation (DBO), Multi-Verse Optimisation (MVO), and Enhanced Multi-Verse Optimisation (EMVO) and showed that their approach achieves faster convergence and lower computational complexity. \\

\begin{figure*}
    \centering
    \includegraphics[width=6.5in]{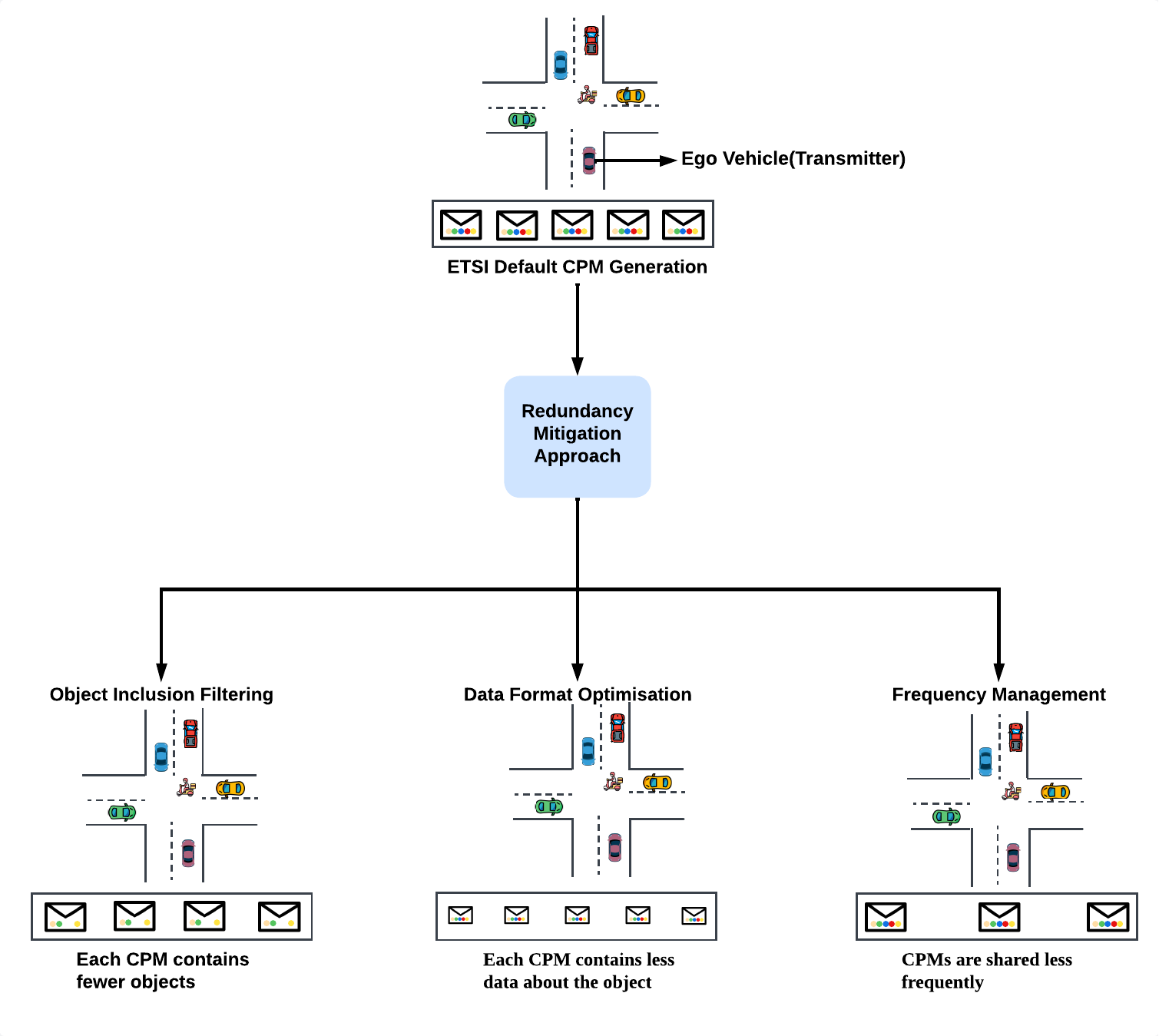}
    \caption{Visualisation of redundancy mitigation taxonomy}
    \label{fig:Illustration_of_RM_Taxonomy}
\end{figure*}

\parindent0pt\subsubsection* {A.1 Short Summary - Object Inclusion Filtering approaches}\hspace*{\fill} \\
Object inclusion filtering attempts to more selectively report only the most relevant objects to reduce CPM overhead. This improves data relevance, reduces channel load, and enhances situational awareness but has drawbacks, such as the potential for missing relevant updates, complexity in implementation, and the need for accurate prediction models. Representative methods include threshold‐based approaches such as \textit{Pre‐Filter}~\cite{thandavarayan2020redundancy} or deep learning/data driven approaches such as DDQN~\cite{ghnaya2023distributed}). Existing schemes face drawbacks: (a) those based on fixed thresholds may not be suitable for all traffic scenarios. This requires further research to fully understand the parameterisation of such thresholds for specific environmental contexts; (b) they often overlook the functional safety implications of an object (e.g. sudden acceleration in the opposite direction may be inconsequential); and (c) AI‐based solutions demand considerable computing resources and may not always fulfil real‐time safety requirements.
\subsection{Data Format Optimisation Approaches}
This category of mitigation techniques aims to reduce the amount of data sent about each object, as illustrated in the middle of Fig. \ref{fig:Illustration_of_RM_Taxonomy}. In contrast to filtering out entire objects, the focus is on compressing or streamlining the object information itself. 

\subsubsection{Dynamic Dissemination Method}
\label{Dynamic Dissemination}
Proposed by Allig et al.~\cite{allig2019dynamic}, this method modifies the data format of objects in the POC by using a complex local fusion module and state estimation process to predict the current and future positions and velocities of the objects. Initially, Bayesian object estimation algorithms, including \textit{constant velocity (CV)}, \textit{constant acceleration (CA)}, \textit{constant turn rate and velocity (CTRV)}, and \textit{constant turn rate and acceleration (CTRA)} models, are used to estimate object states. An \textit{Interacting Multiple Model (IMM)} algorithm then switches between these models based on mode probabilities. The IMM algorithm applies multiple models to account for the changing dynamics of objects, assigning a probability to each model and using these probabilities to predict the object's state more accurately. Instead of fully including an object's distance, angle, GPS, and velocity information, the method retains only the necessary higher-order object attributes based on the predicted patterns. Additionally, consistent inter-trajectory fusion is performed using \textit{covariance intersection (CI)} and its variants to minimise the computational load while ensuring optimal fusion accuracy. The author's evaluation showed that the proposed Dynamic Dissemination method reduced the \textit{Root Mean Squared Error (RMSE)} for position by 25\%, heading by 15\%, and velocity by 20\% when compared to the baseline methods. The baseline used by the authors in \textit{Dynamic Dissemination} is not the ETSI standard CPM but the EPM.

\subsubsection{Coopernaut Method}
\label{Coopernaut}
The \textit{Coopernaut} method~\cite{cui2022coopernaut} proposed by Cui et al., addresses the limitations of line-of-sight perception, data vulnerability in extreme situations, and high CPS bandwidth requirements for autonomous vehicles. The authors developed \textit{AUTOCASTSIM}~\cite{autocast}, a network-enhanced autonomous driving simulation framework, to evaluate the performance of CP in accident-prone scenarios. \textit{Coopernaut} aggregates point cloud data observed by LiDAR and received from surrounding vehicles, using a \textit{Point Transformer} to represent critical detection points efficiently.  The \textit{Point Transformer}, a self-attention network for point cloud processing, learns compact point-based representations by reasoning about non-local interactions among points, producing permutation-invariant representations that are effective in aggregating multi-vehicle point clouds. By filtering out unimportant information and including only objects that impact vehicle operation, the method optimises data transmission. This filtering is done by processing the raw point clouds into keypoints, each associated with a compact representation learned by the Point Transformer blocks, ensuring that only relevant information is included. These keypoints are aggregated and spatially transformed into the ego vehicle's frame, allowing for efficient data sharing among vehicles and improving collective perception performance.

When evaluated, \textit{Coopernaut} demonstrates a 14.5\% improvement in collective perception performance, with \textit{Coopernaut} achieving a success rate of 90.5\% in an Overtaking scenario compared to 81.9\% for Early Fusion and 70.0\% for Voxel GNN~\cite{he2022svga}, a model that processes local point cloud data onboard each vehicle and shares voxel-based representations with the ego vehicle for control, utilising a graph neural network to aggregate these representations for improved driving decisions. Additionally, there was a 90\% reduction in communication cost for mobile nodes, reducing the bandwidth requirement from 60 Mbps if a raw point cloud data was transmitted to just 5.1 Mbps when transmitting key processed data.

\subsubsection{Deep Feature Fusion Method}
Yuan et al. propose an approach~\cite{yuan2022keypoints} that aims to enhance CPS by sharing high-precision data while minimising communication overhead. It refines sensor data from cameras, LiDAR, and GPS to generate CPMs that contain only the most relevant features. The feature extraction process involves using a 3D voxel-based sparse \textit{Convolutional Neural Network (CNN)} backbone to process point cloud data, which is then voxelised, down-sampled, and projected into BEV features. \textit{Farthest Point Sampling (FPS)} is applied to select keypoints, ensuring an even distribution of sparse keypoints. These keypoints are aggregated using the \textit{Voxel Set Abstraction (VSA)} module, which compiles voxel features of varying resolutions. The selected keypoints and their features are split into two paths: one for generating CPMs and the other for correcting localisation errors. The generated proposals and keypoints are then transformed using geometric transformations, clustered using \textit{K-means clustering}, and merged using \textit{Non-Maximum Suppression (NMS)} to ensure accurate detection and minimise false positives. The evaluation of this method showed improvements in both efficiency and accuracy. Detection accuracy was enhanced by approximately 9\%, and the size of the CPMs was reduced to less than 0.3 KB, which is about 50 times smaller than the BEV feature map sharing used in previous studies. 

\subsubsection{Cooperverse Method}
The \textit{Cooperverse} method, proposed by Bai et al.~\cite{bai2023cooperverse} employs random priority filtering and Manhattan Distance calculations to efficiently transmit data in CPMs. \textit{Random Priority Filtering} assigns diverse priorities to detected objects, and \textit{Manhattan Distance} is used to calculate the distance between the detected object and the ITS-S sensor, which differs from the ETSI specified \textit{Distance-based} RMR mechanism summarised in Table~\ref{VoI_Calculation_Methods} and based on Euclidean distance. Feature extraction is then performed using a structured approach where sensor data is voxelised and processed to identify key features. Specifically, features are sifted through a feature deck, prioritised, and selected based on their relevance and object distance to the sensor.

The author's evaluation showed that \textit{Cooperverse} maintains \textit{mean Average Precision (mAP)} classiciation performance until the number of feature cells in the deck decrease to 35,000, resulting in a 14.5\% improvement in \textit{mAP} under an 8.9MB per frame communication constraint. \textit{mAP} is a metric used to evaluate the accuracy of object detection systems, reflecting the system's ability to accurately identify and localise objects within a given dataset correctly. Additionally, substantial bandwidth savings were demonstrated, reducing communication costs by 90\% while only experiencing a minimal 1.7\% reduction in \textit{mAP}. 

\subsubsection{Relayed CPS}
Huang et al.~\cite{huang2023relayed} propose a relay based version of the CPS framework to extend the perception range of vehicles beyond one-hop neighbours using an append-and-forward relay mechanism. The framework modifies the standard structure of CPMs by introducing new fields such as “originatingStationID” and “objectAge” to handle the relayed data effectively. Additionally, the author introduces a redundancy mitigation method using a counter-based mechanism to prevent network congestion caused by repeatedly relaying the same object information. This method limits redundant object information based on a predefined threshold (K), ensuring that only relevant and fresh objects are included in subsequent CPMs. Furthermore, the paper suggests accumulating delays between relaying stations rather than relying on precise time synchronisation, ensuring the AoI is managed accurately in a dynamic vehicular network.

The authors evaluate the proposed method on a three-lane highway and compared against the \textit{ETSI CPS TR (2019)} approach. The comparison showed that the relayed CPS mechanism improved the average perceived object count and the perception range, achieving over a 200\% enhancement. Despite a slight increase in the AoI, it remained within acceptable limits ($\leq 50 ms$). The CBR increased with higher market penetration rates and larger values of the counter threshold (K). The proposed method showed a significant improvement in the number of objects that vehicles perceive. The authors state that the relay mechanism increased CBR and duplicate object rates, but that the trade-off is acceptable, given the significant improvements in perception capabilities.

\parindent0pt\subsubsection* {B.1 Short Summary - Data Format Optimisation approaches}
Data format optimisation approaches aim to minimise message overhead by minimising the object data included in CPMs. Representative methods include \textit{Coopernaut}~\cite{cui2022coopernaut}, which replaces complex data structures with simpler geometric primitives like points and lines, and \textit{Relayed CPS}~\cite{huang2023relayed}, which avoids redundant forwarding by adding additional fields. However, relying on non‐standard data formats raises concerns about their broader applicability across diverse systems and standards and whether sufficiently accurate/intelligent autonomous vehicle decisions can be made when there is loss of detailed information. It also involves complex data processing and computational intensity.\\

\subsection{Frequency Management Approaches}
This category of mitigation techniques aims to decrease how frequently CPMs are transmitted by adjusting how often CPMs are sent based on traffic density, environmental complexity, or network conditions, as illustrated in the bottom-right of Fig. \ref{fig:Illustration_of_RM_Taxonomy}.

\subsubsection{Value-Anticipating Method}
Proposed by Higuchi et al.~\cite{higuchi2019value}, this method aims to predict the VoI of objects based on perceived usefulness to the recipient. The concept of this method is consistent with the Utility-based RMR mechanism proposed by ETSI, as described in Table.~\ref{VoI_Calculation_Methods}. \textit{Kalman filters} are used to predict object states, thereby providing an estimation framework for the information's value. Information entropy quantifies the value of the information, helping to decide which data is essential and calculating this by assessing the reduction in uncertainty about an object's state when new information is received, thus determining its importance. Additionally, \textit{Probabilistic Models} estimate message reception under varying network conditions to determine the likelihood of successful transmission. Under congested network conditions, transmission control mechanisms may delay or cancel the transmission of less valuable messages, ensuring that more critical data is transmitted reliably. The performance evaluation of this method was compared against the default POC inclusion rules. It showed an improved packet reception rate of approximately 25\%, assuming a 50\% penetration rate, although at the expense of a slight increase of 1.3m in object tracking error.


\subsubsection{When2com Method}
\label{When2com}
Liu et al. proposed the \textit{When2com} method~\cite{liu2020when2com} which proposes clustering ITS-S' into communication groups and determines the optimal transmission frequency of cooperative perception data within a given group. ITS-S' generate \textit{compact query vectors} and \textit{larger key vectors}. \textit{Compact query vectors} are low-dimensional representations of the ITS-S' local observations (non-ETSI compliant), designed to minimise bandwidth usage when broadcast. \textit{Larger key vectors} are more detailed representations retained locally, used to assess the relevance of incoming queries from other agents. \textit{When2com} proposes using a handshake communication mechanism involving three stages: \textit{Request}, \textit{Match}, and \textit{Select}. The purpose of \textit{Request} is for each ITS-S to express its need for information by broadcasting a compact query vector based on its local observations. Each receiving ITS-S will invoke the \textit{Match} stage which involves computing similarity scores between the requesting ITS-S' compact query vector and it's own \textit{larger key vector} to identify potential communication partners. This score is calculated using a learned similarity function, often a dot product or another similarity measure, which quantifies how relevant the receiving ITS-S information is to the requesting ITS-S. The final phase, \textit{Select}, is done by choosing the communication partners with the highest matching scores thereby forming a communication group. Perception data is shared cooperatively within the communication group. This method also includes \textit{Self-Attention} mechanisms, which allow agents to decide when communication is necessary based on the sufficiency of their local information. Two variant of the scheme are proposed: \textit{CatAll} which is a naive, fully connected model that concatenates all ITS-S’ encoded image features before subsequent processing stages, whereas \textit{Auxiliary‐View Attention} employs the \textit{Self-Attention} mechanisms to weight supporting ITS-S’ views. \textit{Who2com} is a previous approach by the same authors, optimising partner selection based on current observations but focusing more on inter‐robot rather than V2X communication.

When evaluated, \textit{When2com} shows some performance gains. Under a single-request, multiple‐support scenario, where one vehicle's data may be corrupted but a clean version exists among neighbouring vehicles, it lowered bandwidth usage by approximately 51\% compared to the \textit{Who2com} baseline, from 2MB per frame (MBpf) to 0.98MBpf. In the more complex multiple‐request multiple‐support scenario, where multiple vehicles may have corrupted data and all require accurate information, it achieved a 84.6\% reduction in overhead relative to fully connected methods such as \textit{CatAll} and \textit{Auxiliary‐View Attention}, and a 23\% reduction compared to \textit{Who2com}, reducing the bandwidth from 2.5MBpf to 0.385MBpf.

\subsubsection{P-Consistence Method}
The authors in~\cite{huang2020data} propose an approach where an ITS-S, using its own local sensor readings as well as perception data shared cooperatively by other vehicles, calculates a transmission probability for its tracked objects based on those less likely to be detected by other CAVs. It then adjusts the transmission frequency at which the cooperative messages are sent based on this. The proposed scheme also adjusts the transmission probability based on different traffic densities and CAV penetration rates to balance the trade-off between communication overhead and system reliability. 

\textit{P-Consistence} is compared with three baseline approaches: \textit{Default ETSI POC Inclusion Rule}, where CAVs transmit all tracked objects ensuring maximum situational awareness; the \textit{Fixed Probability (Fixed-p) Scheme}, where CAVs transmit their tracked objects with a predefined probability (e.g., 50\% or 60\%), reducing communication overhead; and the \textit{Modified Irresponsible Forwarding (IF) Scheme}, which prioritises transmissions from CAVs farther from the tracked objects to maximise coverage. Simulation results demonstrated that \textit{P-Consistence} could reduce communication overhead in medium to dense traffic conditions while maintaining the desired share ratio of sensory information. Compared to the \textit{Default ETSI POC Inclusion Rules}, which result in an average payload size of around 30 tracked objects per message, the \textit{P-Consistence} method reduces the average payload size to about 12 tracked objects per message while achieving similar situational awareness. It also outperforms the \textit{Fixed-p} scheme by adaptively adjusting transmission probabilities based on local density estimates, maintaining a consistent share ratio of 95\% compared to the variable 85-95\% achieved by the fixed-p scheme. Furthermore, in V2V based collective perceptions, the \textit{P-Consistence} method surpasses the modified \textit{IF} scheme by achieving comparable V2V awareness (around 92\% of objects shared within the V2V range) with significantly reduced communication overhead, reducing the payload size by approximately 30\% compared to the \textit{IF} scheme.

\subsubsection{Deep Q-Learning Method}
Abdel-Aziz et al. present a \textit{Deep Q-Learning} method~\cite{abdel2021v2v} uses a Branching Dueling Q-Network (BDQ), a type of Reinforcement Learning model, to assess the quality and timeliness of sensory data to adjust the frequency of CPM delivery. Specifically, it tackles the challenges of vehicle association, resource block (RB) allocation, and content selection for CPM. The goal is to maximise vehicular satisfaction in terms of the quality and timeliness of the received sensory information. The proposed method involves two RL problems: one at the RSU level and another at the vehicle level. At the RSU level, the RL agent learns to optimally associate vehicles and allocate RBs. At the vehicle level, each vehicle acts as an RL agent to decide which quadtree-compressed blocks of sensory data to transmit to its paired vehicle. The method uses a BDQ to manage the high-dimensional action spaces inherent in these RL problems, efficiently distributing the representation of actions across network branches while maintaining a shared state representation. The RL agents adjust the transmission frequency of CPMs by evaluating the quality and timeliness of sensor data, quantified by the age of information (AoI) and the probability of RB occupancy, and considering the vehicle's interest in the data based on relevance to its current and future positions and direction. When evaluated, \textit{Deep Q-Learning} enhances vehicle satisfaction, defined as the quality and relevance of received sensor information. Specifically, the BDQ agent maintained robust performance with a large action space of approximately 2 million actions, whereas the Deep Q-Network (DQN) agent did not. The RSU agent's reward increased with the number of training episodes, indicating improved vehicle association, RB allocation, and CPM content selection. Furthermore, the complementary cumulative distribution function (CDF) of vehicular rewards confirmed that trained agents achieved superior results.

\subsubsection{Evidential Deep Learning (EDL) Method}
\label{EDL}
The \textit{EDL} method~\cite{yuan2023generating} is proposed by Yuan et al. who are the same authors that proposed the \textit{Deep Feature Fusion} method~\cite{yuan2022keypoints} already summarised as a data format optimisation approach. While \textit{Deep Feature Fusion} focuses on optimising how objects are represented in CPMs, the \textit{EDL} method modifies when CPMs are transmitted.

The method modifies CPM generation rules by employing a \textit{Convolutional Neural Network Architecture (U-Net)} to process point cloud data from sensors such as cameras, LiDAR, and GPS. The \textit{U-Net-based} backbone learns and aggregates deep features of different resolutions from input point clouds, and these features are used to create BEV maps that help in uncertainty-based CPM selection. This network extracts \textit{Deep Features} and compresses them into 2D features for BEV maps. The \textit{deep features} include point-wise spatial Gaussian distributions, allowing the network to infer the likelihood of points belonging to specific classes (e.g., road, vehicle) and the associated uncertainty. These maps quantify perception uncertainty, highlighting areas of high and low confidence. The approach prioritises areas with low confidence by requesting surrounding ITS-S to send information about these regions. The ego CAV generates a binary request mask by thresholding its uncertainty map, identifying areas of high uncertainty. It then requests additional information from surrounding ITS-S, which responds with masked evidence maps from low-uncertainty areas, ensuring efficient data sharing and reducing redundancy. 

The \textit{EDL} method was compared against several state-of-the-art models, including its own previous \textit{Deep Feature Fusion} method from 2022. The comparative analysis included methods like \textit{CoBEVT, AttFuse, V2VNet, DiscoNet, and Fcooper}, which represent diverse approaches to collective perception, from \textit{Feature Fusion} to \textit{Transformer-based} methods. \textit{V2VNet}, \textit{DiscoNet}, and \textit{Fcooper}, which use graph neural networks and sparse feature maps to enhance perception while reducing communication overhead, are among the \textit{Feature Fusion} methods. \textit{CoBEVT} and \textit{AttFuse}, utilising attention mechanisms and sparse transformers to improve data integration and interpretation, are among the \textit{Transformer-based} methods. 

The results indicated that the \textit{EDL} method outperformed the \textit{Deep Feature Fusion} method, achieving a 54.8\% increase in IoU and a 93\% reduction in CPM size. EDL demonstrated higher dynamic object recognition accuracy than the five methods mentioned in both the OPV2V and V2V4Real benchmarks, with a maximum improvement of about 20\%. This demonstrates a superior trade-off between communication load and perceptual accuracy by leveraging uncertainty information to prioritise data sharing. While the reduction in CPM size is significant, the slight 4\% drop in IoU suggests that the method, although efficient in data reduction, may occasionally exclude critical information necessary for certain driving scenarios.

\subsubsection{STC Method -- TC}
\label{TC}
\textit{Temporal Clustering (TC)}, part of the STC method proposed by Hakim et al.~\cite{hakim2024stc}, focuses on clustering objects based on temporal factors to optimise the frequency of data transmission. The \textit{TC} method specifically addresses the timing of data updates to reduce redundant transmissions and enhance the relevance of the shared information. \textit{TC} predicts the dynamics of objects for the next 1 second in 0.1 second intervals. Each time step is classified as either 'avoidable' (does not meet the threshold and can be left out of the POC) or 'unavoidable' (meets the threshold and must be added to the POC). By doing this, \textit{TC} ensures that only significant changes in object states are transmitted. The information on avoidable time steps is merged into the nearest unavoidable time step, and they are sent together. This reduces the frequency of message sending while ensuring the comprehensiveness and relevance of the data.

The results of applying the \textit{TC} method, combined with \textit{SC} as the \textit{STC} method, showed improvements in communication performance compared to the ETSI standards and other baseline methods such as \textit{LA} and \textit{PF} etc. According to the evaluations, the \textit{STC} method achieved a 41\% reduction in communication payload and increased perception accuracy by up to 37\% compared to the baseline ETSI method. Under low-density scenarios, \textit{STC} increased the reception of new information, showing a 7.9-fold increase in high-density urban environments and a 10.9-fold increase in high-density highway scenarios compared to ETSI. \\

\parindent0pt\subsubsection* {C.1 Short Summary - Frequency Management approaches}
Frequency management approaches concentrate on dynamically adjusting how often CPMs are transmitted. Representative methods include \textit{Value‐Anticipating}~\cite{higuchi2019value}, which decides whether to send the current CPM by predicting if the recipient already perceives the object, and \textit{When2com}~\cite{liu2020when2com}, which employs AI to identify the optimal frequency of communication. However these solutions face key challenges: (a) prediction accuracy heavily influences overall performance, and (b) AI‐based methods can demand significant computing resources, potentially limiting real‐time applicability. As such, while frequency management can reduces unnecessary transmissions, it depends on accurate request mechanisms, faces complexity in sensor fusion, and can result in potential computational overhead.

\subsection{Infrastructure-Based Redundancy Mitigation Mechanisms}
A subset of the afore-mentioned schemes utilise RSUs~\cite{chtourou2021context, huang2023relayed, abdel2021v2v}. However in all cases individual vehicles still invoke some form of redundnacy mitigation to limit the perception information that they cooperatively. In contrast, the method proposed by Singh et al.~\cite{singh2024optimizing} proposes an alternative centralised approach where ITS-S' send their perception updates to an RSU which aggregates this information to broadcast an aggregated CPM. It works as follows: The RSU receives all CPMs transmitted by vehicles within its communication range. It processes the CPMs to extract relevant information on objects and vehicle positions, removing outdated or redundant data. It then calculates global object coverage and uses a sub-modular function to model the vehicle selection problem, where the goal is to maximise object coverage using a minimal subset of vehicles. The RSU employs a greedy algorithm to iteratively add vehicles to the selection set based on their marginal gain in coverage until the desired coverage is achieved or adding more vehicles no longer provides a significant benefit. The selected vehicles then transmit their CPMs in the next time slot, while the non-selected vehicles remain silent to avoid congestion. The RSU dynamically updates the vehicle selection set over multiple time slots, adjusting parameters based on changes in vehicle positions and network conditions to ensure consistent and optimal CPM dissemination. The proposed method is compared with the Minimum Cost Coverage Problem (MCCP) and the standard CPM transmission approach defined in \textit{ETSI CPS TS (2023)}. It is worth mentioning that the author assumes that all vehicles within the range have the ability to generate and send CPM. It outperforms both methods, achieving up to 94.92\% reduction in network load compared to the standard approach and using fewer vehicles to achieve nearly the same object coverage as MCCP, thereby reducing network congestion and communication overhead while maintaining high coverage levels.

\section{Quantitative Simulation Environments} \label{eval}

Research on redundancy mitigation, and Cooperative Perception Systems more widely, spans multiple domains including sensing, communications and environmental/traffic flow dynamics. This reflects the broad range of research questions being addressed in the field. As such, the choice of simulation environment is often determined by the main focus of the research study at the expense of the fidelity of other impacted system component models. Additionally, the complexity of CPS research typically necessitates the adoption of a co-simulation environment, where multiple simulators are integrated to capture the interactions between various subsystems. While co-simulation provides a more comprehensive analysis, it also introduces additional complexity, requiring careful synchronisation and calibration of the different simulation tools to ensure accurate and reliable results. We now review the co-simulation environments used for redundancy mitigation studies within academic literature and aim to help the reader assess which co-simulation environment should be used depending on the focus of a particular research study.

\begin{itemize}
    \item \textbf{Integrated Co-simulation:} Fully integrated co-simulation environments that include detailed models for mobility, networking \& communications, and sensors. Examples include frameworks like AUTOCASTSIM~\cite{autocast} and OpenCDA~\cite{xu2021opencda}, which couple SUMO~\cite{krajzewicz2010traffic} for mobility simulation, NS-3~\cite{henderson2008network} or OMNeT++~\cite{varga2010omnet++} for network simulation, and CARLA~\cite{dosovitskiy2017carla} for sensor modelling to provide a comprehensive co-simulation environment. Representative studies leveraging this approach include \textit{Coopernaut}~\cite{cui2022coopernaut}, which employs \textit{AUTOCASTSIM} to assess cooperative perception strategies.
    \item \textbf{Sensor Abstraction:} Implements thorough network and mobility models but abstracts and simplifies the sensor simulation. Commonly used simulators in this category include SUMO~\cite{krajzewicz2010traffic} for mobility and NS-3~\cite{henderson2008network} or OMNeT++\cite{varga2010omnet++} for network simulation. Frameworks like Veins~\cite{sommer2019veins} and Artery~\cite{riebl2015artery}, which integrate SUMO~\cite{krajzewicz2010traffic} and OMNeT++~\cite{varga2010omnet++}, enhance simulations by combining traffic and network dynamics. Several works have adopted this approach, including \textit{AICP}~\cite{zhou2022aicp}, \textit{VILE}~\cite{lobato2023redundancy}, and \textit{Context-Aware Selection}~\cite{chtourou2021context}, all of which utilise Veins to evaluate their respective redundancy mitigation techniques.
    \item \textbf{Mobility Abstraction:} Implements thorough network and sensor models but abstracts and simplifies the mobility simulation. Simulators in this category often emulate sensor data detected at specific moments and optimise the data at those instances without considering continuous data exchange over time. CARLA~\cite{dosovitskiy2017carla} is often used to model sensors. Either OMNet++~\cite{varga2010omnet++} or an analytical model is used to simulate the network part. This type of simulation setup is often used when the focus of the study is on the impact of the size of the CPM on the network performance. An example from literature that utilises this approach is the \textit{Feature-Based}~\cite{masuda2022feature} method, as it is assessing the influence of vehicle feature recognition on redundancy mitigation.
    \item \textbf{Network Abstraction:} The converse of the \textit{Mobility Abstraction}; it models mobility and sensor aspects in detail but abstracts and simplifies the communication network. Self-driving simulators like CARLA~\cite{dosovitskiy2017carla} are employed in this category, providing realistic environments for testing advanced sensor fusion and perception algorithms. Representative studies following this approach include \textit{Where2comm}~\cite{hu2022where2comm} and \textit{EDL}~\cite{yuan2023generating}, both of which leverage CARLA to examine cooperative perception without direct network simulation constraints.
\end{itemize}

An important finding from the literature is that realistic sensors can only achieve 42\% of the object detection rate of idealised sensors~\cite{li2023study}. Additionally, Andreani et al.~\cite{andreani2023statistical} pointed out that more realistic object detection impacts the size of CPMs. Thus, the choice of simulation environment is highly influential when evaluating the communication network challenges posed by the absence of redundancy mitigation techniques and the accuracy/improvements offered by proposed solutions. Careful consideration enhances the validity and applicability of proposed research findings.

\section{Open research opportunities} \label{orcs}

Several open research challenges remain to ensure that redundancy mitigation techniques can adequately reduce communication load while maintaining adequate perceptual accuracy. This gives rise to a number of outstanding challenges as well as research opportunities:
\subsection{Implications of Redundancy Mitigation Across Domains}
While redundancy mitigation measures may improve the challenges posed to the communication network, ideally with negligible impact on the perception accuracy of the vehicle, it may have unintended consequences for other vehicular and autonomous driving technologies that can benefit from redundant information, e.g. cyber-security, local dynamic mapping, and trust frameworks for shared information. Open research challenges include understanding the impact of redundancy mitigation on these interconnected domains and exploring the trade-off between satisfying the need to reduce redundant communications without compromising the integrity of other vehicular and autonomous driving sub-systems. The over-arching outstanding research challenge is how to identify and maintain the 'ideal' level of redundancy.

\begin{figure}
        \centering
        \includegraphics[width=\columnwidth]{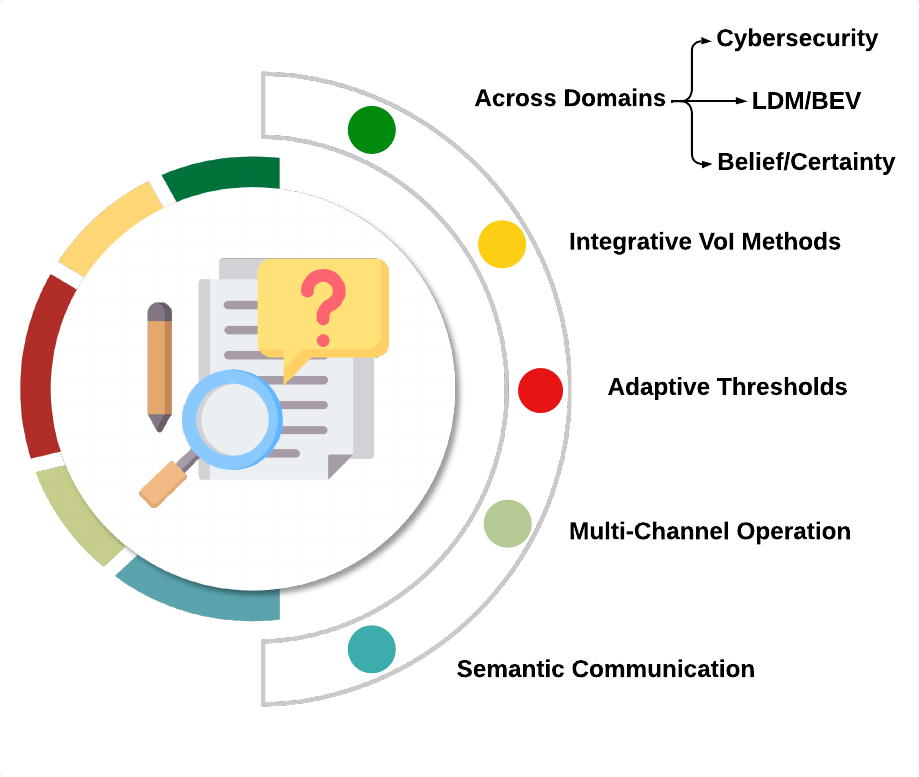}
        \caption{Open Research Challenges in Enhancing Redundancy Mitigation.}
        \label{fig:Frequency-Based RMR}
\end{figure}

\subsubsection{RM and Cybersecurity}
As stated, while redundant messages can be problematic from the perspective of communication network performance, such redundancy may be beneficial for cross-checking the validity of cooperatively shared sensed objects. In particular, redundancy may enable improved detection of false or malicious behaviour through redundant data from honest actors, thereby enhancing the security of V2X communications. Ansari et al.~\cite{ansari2021v2x} reviewed the format of CPMs as outlined in \textit{ETSI CPM TR (2019)} and discussed the standard’s limitations, particularly in terms of security vulnerabilities and the absence of misbehaviour detection mechanisms. The authors conducted a comprehensive threat assessment of \textit{ETSI CPS TR (2019)} using a matrix-based approach, through which they identified 16 potential attacks and acknowledged that redundancy could assist in identifying and potentially eliminating such attacks. These include sensor spoofing, where false sensor data is reported; perceived object spoofing, which involves falsifying object information such as location and velocity; and sensor blinding, where attackers disable sensors, significantly reducing vehicle awareness. Some researchers have investigated the impact of CPM redundancy on cybersecurity~\cite{hadded2019augmented, ambrosin2019design, hadded2020security, ali2023distributed} in specific contexts such as cross-verification of object data and mitigation of sensor blinding etc. Although these studies address specific threats in well‐defined scenarios, there remains significant research opportunities. Thus, a valuable future research direction is to investigate reconciliation methods that balance suitable levels of purposeful redundancy while managing efficient channel usage, thus ensuring that the CPS remains both secure and scalable.



\subsubsection{RM and LDM/BEV}
A \textit{Local Dynamic Map (LDM)} is a four-layer database structure for managing a C-ITS map and real-time vehicle data, enabling vehicles and infrastructure to share information. A \textit{Bird's Eye View (BEV)} map is a geometric representation of the scene from above the vehicle, fusing multiple sensor data to provide an integrated view. While LDMs use database queries to find objects in the surrounding area, BEVs rely on image recognition algorithms for environmental perception. Both LDMs and BEVs are representations of the surrounding environment and are vital in assisting autonomous vehicles in making decisions. Separately, Kalman filters are widely used in autonomous driving for sensor fusion and state estimation~\cite{veysiimplementation}. For vehicles, Kalman filters help integrate sensor data to estimate position, speed, and direction accurately, which facilitates precise navigation and control. For perceived objects, Kalman filters are effective for accurate tracking. Researchers in~\cite{taddei2024multi} and~\cite{cui2022coopernaut} propose using CPMs to enhance LDMs or BEVs to create a more comprehensive and accurate environmental representation and for estimating the future trajectory of an object.

However, to make accurate predictions and ensure safe driving, the Kalman filter needs as much information as possible about the tracked object. While this information may be included in CPMs, it is largely captured in the optional CPM fields. As pointed out by Xhoxhi et al.~\cite{xhoxhi2023first}, adding these optional fields inevitably increases the size of the CPM, which in turn affects the efficiency of network communication. Therefore, a key research question is how to strike a balance between adding sufficient optional data for effective Kalman filter prediction and minimising network load. Deciding which fields to include and which to exclude in order to optimise both objectives remains an outstanding challenge.

\subsubsection{RM and Belief/Certainty}
When vehicles cooperatively share information, it may be desirable for the receiver to derive a collective belief/confidence in the shared data relating to a single object. This could be to identify inaccurate data that has been shared by a malign actor but also data that may be unintentionally inaccurate or display low confidence due to onboard sensor limitations in a given environment. Deriving such belief/confidence is particularly important when contradictory data is cooperatively shared i.e. a fast cyclist vs a slow-moving motorbike. Redundant information supports trust algorithms/frameworks by enabling cross-validation of data from different sources, which can improve the accuracy of shared information and the confidence in the reliability of perceived objects. Trust management must consider various aspects of data quality, including sensor accuracy, time delays, and potential malicious behaviour. Redundant information can help filter out unreliable information, but only if managed correctly. Redundancy also increases the complexity of deciding which data to trust, especially in dynamic environments. A valuable future research direction is to explore how to balance the benefits of redundancy for trust and certainty with the challenges of communication overhead and computational complexity. This requires developing adaptive trust management mechanisms that dynamically adjust the level of redundancy based on source reliability, scenario criticality, and network conditions to ensure accurate decision-making by CP systems.
\subsection{Integrative VoI Methods \& their impact on control decisions}
\textit{ETSI CPS TS (2023)} has recommended that using multiple VoI methods together can improve overall redundancy efficiency. However, it is not clear which redundancy mitigation techniques should be used together and in what order. In~\cite{delooz2022analysis}, Delooz et al. suggested that the \textit{Self-Announcement-Based} method should not be used as a stand-alone rule but rather as a supplement to another method and noted that the \textit{Frequency-Based} method requires additional mechanisms to prevent multiple vehicles from sending the same object information simultaneously. 
Recently, and as discussed in Section~\ref{EvaluationStandardisedRMR}, Sakr~\cite{sakr2024evaluation} analysed the coupling of \textit{Self-Announcement-Based} and \textit{Dynamic-Based} methods as well as \textit{Frequency-Based} and \textit{Dynamic-Based} techniques and Malik et al.~\cite{Malik2024-bt} evaluated the coupling of \textit{Self-Announcement-Based} and \textit{Distance-Based} techniques. In both cases, the evaluations showed that the combination of schemes performs worse than if a singular mitigation method is used. An analysis as to why these combinations performed worse, i.e. whether they contradict each other, was not provided. As observed by Thandavarayan et al., in Section~\ref{eRMLA}, in some cases, mitigation methods may actively work against each other, e.g. Look Ahead and Pre-Filter~\cite{thandavarayan2023scalable}, unless actively designed otherwise.


Therefore, several problems remain. Firstly, existing research has not evaluated all eight VoI methods proposed in \textit{ETSI CPS TS (2023)}. Secondly, there is not a sufficient study into which mitigation methods are complimentary (or conversely contradictory) and in which sequence they should be invoked. Finally, even when VoI is calculated, its application to vehicle control has not been adequately considered. Authors in~\cite{higuchi2019value} are the only to date to have attempted to approach redundancy mitigation by predicting the VoI based on perceived usefulness to the receiver and linked this to potential vehicle control decisions. Thus, open challenges remain to devise combined VoI calculation methods that can incorporate factors beyond network communication, such as vehicle control and safety considerations.

\subsection{Adaptive Tuning of VoI Redundancy Thresholds}
Each of the eight VoI-based methods for redundancy mitigation proposed in \textit{ETSI CPS TS (2023)} utilise fixed thresholds to determine their behaviour. For instance, the \textit{Frequency-based} method employs fixed values like \textit{N\_Redundancy (1, 3, 5)}, while the \textit{Dynamic-based} method has fixed thresholds such as \textit{P\_Redundancy (8m)} and \textit{S\_Redundancy (1m/s)}, and the \textit{Angle-based} method has a fixed threshold \textit{A\_Redundancy (8$^{\circ}$)}. These static thresholds are unable to adapt to varying environmental and dynamically changing conditions, potentially causing inefficiencies, e.g. when utilising the \textit{Dynamic-Based} method in a highway environment, the position of the vehicle changes quickly, while the speed may remain the same. Using fixed thresholds may result in many redundant CPMs, and ideally, a low \textit{S\_Redundancy} and a high \textit{P\_redundancy} should be used. In a congested urban environment, the vehicle moves slowly, with the position changing very little, but with frequent speed fluctuations due to stop-and-go traffic, and this may also result in redundant CPMs. Ideally, a high \textit{S\_Redundancy} and a low \textit{P\_redundancy} should be used. See Figure~\ref{fig:Dynamic-Based RMR} and Section~\ref{EvaluationStandardisedRMR} for further details. 

Therefore, the development of a context-aware mechanism to dynamically tune thresholds is necessary. Such a mechanism would ensure context awareness while minimising redundancy. Previous research~\cite{chtourou2021context} attempted to use network metrics like CBR for dynamic threshold selection. However, more advanced techniques, such as reinforcement learning, could provide more accurate and effective solutions for adaptive threshold tuning.
\subsection{Leveraging Multi-Channel Operation for Efficient Redundancy Control}
As outlined in Section~\ref{network}, the \textit{ETSI CPS TS (2023)} proposes leveraging Multi-Channel Operation (MCO) to prioritise high VoI objects that are critical to vehicle situational awareness by transmitting them on the preferred channel. However, due to the absence of a single predominant V2X communication technology, the practical implementation of this approach remains uncertain.

ITS-G5, based on \textit{IEEE 802.11p}, has traditionally utilised the entire allocated spectrum in a single-channel operation but was recently enhanced to support MCO through amendments in the ETSI standards as part of C-ITS Release 2. Key standards introducing MCO for ITS-G5 include:~\cite{etsiintelligentMCOArchitecture}, which extends the C-ITS communication architecture to accommodate MCO; ~\cite{etsiintelligentMCOStudy}, which defines the MCO architecture, entities, and functionalities; ~\cite{etsiintelligent2022}, which covers MCO functionalities at the facilities layer; ~\cite{etsiintelligentMCOAccessLayer}, which details MCO at the access layer. These enhancements enable ITS-G5 to support features such as application-driven prioritisation, adaptive channel switching, and load balancing across multiple channels. MCO allows ITS-G5 to dynamically assign channels to messages based on their priority and current channel conditions, facilitating better spectrum utilisation and reduced latency for safety-critical messages. Conversely, C-V2X, defined in 3GPP Releases 14 and 15, utilizes Single-Carrier Frequency Division Multiple Access (SC-FDMA) in the physical layer, supporting flexible configurations for sub-channel bandwidth allocation~\cite{3GPP_TR_21.914, 3GPP_TS_22.186}. C-V2X sub-divides frequency into sub-channels using different Resource Blocks (RBs). This subdivision function is similar to MCO in that it allows different data streams to be transmitted simultaneously on different frequencies as distinct sub-channels. C-V2X's sub-channel approach, combined with Semi-Persistent Scheduling (SPS), provides some of the benefits of MCO, particularly in terms of maintaining a structured communications environment. 

Defining practical methods for implementing VoI-based channel selection in both ITS-G5 and C-V2X is required. A critical research question is how to develop channel selection algorithms that effectively assign VoI-based messages to preferred channels/sub-channels to optimise communication efficiency and prioritise delivery of high VoI packets. Although the \textit{ETSI CPS TS (2023)} recommends this approach for ITS-G5, it remains a theoretical proposal without practical implementation. In C-V2X, VoI considerations in channel selection or prioritisation mechanisms for high-priority packets are currently unaddressed. Developing suitable mechanisms to identify high VoI packets and consistent channel selection algorithms that consider dynamic network conditions, channel occupancy, and data priority are necessary. Such channel selection algorithms should also prevent erratic behaviour where vehicles alternate frequently to the same preferable channel. Similar to congestion control algorithms, it is imperative to apply consistent, rather than reactive, algorithms across all vehicles to ensure stable and efficient network operation.

\subsection{Semantic Communication for Efficient Data Transmission}
The current CPM format includes optional fields for a detailed description of an object. As outlined in Section~\ref{NetworkChallengesPosedByCP}, these can significantly increase the size of the CPM. In some driving scenarios, it may not be necessary to transmit detailed data about detected objects. Thus, if the main cause of network congestion is the size of the CPM, filtering out unnecessary information becomes particularly important. Semantic communication (SEM-COM)~\cite{yang2022semantic} offers a promising solution by transmitting not the detailed properties of objects but rather high-level semantic information/actions, such as fields indicating whether to slow down, stop, or proceed with caution. By reducing the size of the CPM while retaining the information necessary for decision-making, SEM-COM can improve network efficiency, particularly in resource-constrained environments. Recent research has demonstrated an emerging interest in employing SEM-COM to enhance the efficiency of CP. For instance, approaches such as \textit{Where2comm}~\cite{hu2022where2comm} and \textit{SwissCheese}~\cite{xie2024swisscheese}  utilise advanced AI algorithms to semantically extract sensor data, transmitting compact semantic information instead of standard CPMs. While these approaches do not explicitly claim to target network load reduction by mitigating redundancy, they provide a potential direction for advancing CP systems.

However, SEM-COM also presents several challenges that require further research. Firstly, the usability and efficiency of SEM-COM rely heavily on vehicles having the same background knowledge to decode the high-level semantic information accurately. If a vehicle lacks this shared context, it may misinterpret the received information. Secondly, the process of converting detailed data into high-level semantic information requires significant computing resources and energy, which could lead to delays in message generation. Finally, SEM-COM typically involves processing large amounts of context-rich, privacy-sensitive information. Developing effective methods for desensitizing this data while ensuring privacy remains an open challenge.

\section{Conclusion}
This paper critically reviews redundancy mitigation strategies for the vehicular collective perception service. It provides a comparative analysis of ETSI proposals, including the \textit{ETSI CPS TR (2019)} and the \textit{ETSI CPS TS (2023)}, highlighting the shift toward VoI-based measures for object inclusion and frequency management. A novel taxonomy is proposed, categorising ETSI specifications and academic literature according to the approach taken i.e., \textit{object inclusion filtering}, \textit{data format optimisation}, and \textit{frequency management} and critically reviewing said strategies. The analysis highlights the dependency of redundancy mitigation performance on parametrisation and contextual environmental factors, such as traffic density and object dynamism. The paper further underscores the unexplored areas within redundancy mitigation, including multi-channel operation for prioritising high VoI messages, the integration of semantic communication frameworks, the "ideal" level of redundancy and the impact of real-world perception accuracy on communication requirements, offering opportunities for research questions.

\section{Acknowledgements}
This publication has emanated from research supported in part by a grant from Science Foundation Ireland under Grant number 18/CRT/6222. For the purpose of Open Access, the author has applied a CC BY public copyright licence to any Author Accepted Manuscript version arising from this submission.

\bibliography{references}
\bibliographystyle{IEEEtran}
\end{document}